\documentclass[11pt]{article}

\usepackage[dvipsnames]{xcolor}
\usepackage[utf8]{inputenc}
\usepackage[margin=2.5cm]{geometry}
\usepackage[english]{babel}
\usepackage{authblk}
\usepackage{abstract}

\usepackage{amsmath,amssymb,amsfonts,amsthm}
\usepackage{mathrsfs}
\usepackage{bm}
\usepackage{physics}

\usepackage{graphicx}
\usepackage{xcolor}
\usepackage{abstract}
\usepackage{tikz}
\usetikzlibrary{shapes.misc,decorations.markings,arrows,decorations.pathmorphing,backgrounds,matrix,patterns,positioning,decorations.pathreplacing,calc}

\usepackage{float}
\usepackage{caption}
\usepackage{subcaption}
\usepackage{booktabs,array,multirow,makecell}
\usepackage{enumitem}
\usepackage{comment}
\usepackage{cite}
\usepackage[hyperfootnotes=true,colorlinks=true,linkcolor=blue,citecolor=purple]{hyperref}

\graphicspath{{./figures/}}
\numberwithin{equation}{section}
\newcolumntype{M}{>{$}c<{$}}

\definecolor{dgreen}{rgb}{0,0.5,0}
\definecolor{darkblue}{rgb}{0,0,0.6}
\definecolor{darkgreen}{rgb}{0,0.5,0}
\definecolor{lblue}{rgb}{0.5,0.5,1.0}
\definecolor{purple}{rgb}{0.4,.2,0.7}
\definecolor{ggray}{rgb}{0.85,.85,0.85}
\definecolor{dpink}{rgb}{1,0.5,0.5}

\newcommand{\bes}{\begin{equation}\begin{split}}
\newcommand{\ees}{\end{split}\end{equation}}

\renewcommand{\i}{\mathrm{i}}

\newcommand{\lp}{\left(}
\newcommand{\rp}{\right)}

\newcommand{\R}{\mathbb{R}}
\renewcommand{\O}{\mathcal{O}}
\newcommand{\M}{\mathcal{M}}

\newcommand{\pd}{\partial}
\newcommand{\evs}[1]{\langle #1 \rangle} %

\newcommand{\rt}{\sqrt{2}}
\newcommand{\hf}{\tfrac{1}{2}}

\begin{document}

\title{\LARGE\bf 
\vspace{1cm}

Solving the SU(2) BFSS model
} 
\author{Kabir Bajaj and Henry W. Lin} %
\affil{Joseph Henry Laboratories and Leinweber Forum for Theoretical Physics,\\
    Princeton University, Princeton, NJ 08544, USA}
\date{}

\thispagestyle{empty}
\maketitle
\begin{abstract}
We numerically solve the ground state wavefunction of SU(2) BFSS matrix quantum mechanics. We compare ground correlators against islands obtained using the matrix bootstrap and find agreement up to five significant figures. The power law tail of this wavefunction determines the three-point amplitude, which agrees with the computation of Herderschee and Maldacena. Beyond the ground state, we extract the non-perturbative four-point scattering amplitude using L\"uscher's method. In the appropriate regime, we find agreement with the Born-Oppenheimer approximation that yields the well-known $v^4/r^7$ interaction. We also find explicit bound states with positive energies in various SO(9) irreps. These are closely related to the bound states whose existence was recently discovered by OpenAI. Finally, we identify metastable resonances in the spectrum.

\end{abstract}

\vspace{1in}

\pagebreak

\setcounter{tocdepth}{3}

{\setlength{\parskip}{0pt}\tableofcontents}
\newpage
\section{Introduction} \label{sec:bfss}

The Banks-Fischler-Shenker-Susskind (BFSS) matrix model~\cite{Banks:1996vh} is a quantum mechanical system that arises as the low energy description of D0-branes or the dimensional reduction of $\mathcal{N}=4$ SYM with SU($N$) gauge group \cite{deWit:1988wri,Witten:1995im}. 
Solving it at large $N$ would give a treasure trove of holographic data on strings \cite{Itzhaki:1998dd, Hanada:2008ez}, branes, black holes, and M-theory \cite{Banks:1996vh}. It is therefore high on our wish-list of particular quantum systems to solve.
For recent reviews of this model, see \cite{Maldacena:2023acv, Lin:2025iir, Yin:StringTheory}. 

In this work, we numerically solve various aspects of the BFSS model with gauge group SU(2). 
This is clearly far from even the 't Hooft limit where one expects a 10-dimensional gravity dual \cite{Itzhaki:1998dd}. Nevertheless, even for $N=2$ the model exhibits rich physics. 
In particular, the effective coupling in this model is $\sim g^{2}_\text{YM} / E^3$, so at low energies the model is strongly coupled. 
The ground state provides a natural starting point for exploring this strongly coupled regime; our first order of business is to obtain a high precision numerical solution of the ground state wavefunction. 
While the SU(2) theory is known to have a unique normalizable zero-energy state~\cite{Yi:1997eg,Sethi:1997pa,Moore:1998et,OpenAIMath:2026bfssunique} that is an SO(9) singlet~\cite{Sethi:2000zf}, constructing the explicit, non-perturbative wavefunction has remained an open problem.
Using gauge and SO(9) invariance, we may express the wavefunction $\psi$ as a map from 3 real bosonic coordinates to $\mathbb{C}^d$ where $d=2066$.
The supercharge equations $Q_\alpha \ket\Omega = 0$ then become a system of coupled first-order differential equations in three variables, which we solve numerically by some modified variational method. Solving these equations resolves both the strongly interacting core of the bound state and its asymptotic region, where the two D0-branes are widely separated. 

We also compare expectation values computed from the wavefunction with independent bounds, obtained using the matrix bootstrap method \cite{Lin:2020mme,Han:2020bkb, Lin:2023owt, Lin:2024vvg, Kazakov:2024ool}. 
We checked that various low-level correlation functions lie within the allowed intervals, with the tightest comparisons resolving approximately five significant figures. The agreement for even rather complicated observables such as $\ev{\Tr X^8}$ (see Table~\ref{tab:moments}) greatly increased our confidence in these numerical results. 
As another indication of the fidelity of our numerical results, the energy of our solution is $E \approx 1.6 \times 10^{-9} g_\text{YM}^{2/3}$, close to the supersymmetric value $E=0$.

Next we study scattering amplitudes. An important simplification that happens in the SU(2) model is that the only possible amplitudes are the $1 \to 2$ amplitude (the three-point amplitude computed in \cite{Herderschee:2023pza}) and the $2 \to 2$ amplitude. 
To extract the three-point amplitude, we study the tails of the ground state wavefunction{~\cite{Lin:2014wka}}, that involve splitting the $2\! \times\! 2$ matrices into two blocks of size $1\!\times\!1$ and $1\!\times\! 1$. Since all asymptotic states in the SU(2) model involve 2 D0-branes, $2\to2$ scattering is purely elastic. 
Therefore all of the scattering physics is completely encoded in the phase shift $\delta_\ell(E)$.
We extract these phase shifts by following L\"uscher's method~\cite{Luscher:1990ux}: we place the BFSS model in a large, spherically symmetric box, and compute the energy levels as a function of the radius of the box. The phase can then be extracted from these eigenvalues. A closely related approach was used to extract the S-matrix and metastable resonances in supersymmetric quantum mechanics in~\cite{Balthazar:2016utu}.

We next study normalizable states at positive energy. The existence of infinitely many such states in SU(2) BFSS was recently established using states with large angular momentum~\cite{OpenAIMath:2026bfss270}. We refine this argument to prove the existence of infinitely many states with zero or small amounts of angular momentum. 
These existence arguments do not determine the energies or
wavefunctions of the bound states. We compute both numerically
for states in several SO(9) representations, adapting the methods
used for the ground state and scattering states. The same
calculations also reveal metastable resonances, whose physical
interpretation remains an interesting question for future work.

{\bf Note added:} after we obtained most of these results, we learned of simultaneous work by Michel, Scheinpflug, and Yin \cite{MichelScheinpflugYin:2026} which contains some overlap with our results. We thank them for sharing their manuscript with us. Related work on small bootstrap islands in SU(2) BFSS will be reported in \cite{BajajEtAl:BFSSBootstrap}.
\section{Ground state of SU(2) BFSS}\label{sec:ground}

\subsection{Invariant formalism}

The ground state $\ket{\Omega}$ is defined by $Q_\alpha\ket{\Omega}=0$.
The ground state is invariant under SO(9) rotations and SU(2) gauge transformations generated by $C$.
The sixteen supercharges and their algebra are
\begin{equation}\label{eq:QandC}
\begin{split} 
Q_\alpha&=\Tr\lp P_I\gamma^I_{\alpha\beta}\psi_\beta\rp-\frac{\i}{2}\Tr\lp[X^I,X^J]\gamma^{IJ}_{\alpha\beta}\psi_\beta\rp,\\
\{Q_\alpha,Q_\beta\}&=2\delta_{\alpha\beta}H+2\gamma^I_{\alpha\beta}\Tr\lp X^IC\rp,\qquad
C=-\i[X^I,P^I]-\psi_\alpha\psi_\alpha-\mathbf 1 .
\end{split}
\end{equation}
We work in units where $g_\text{YM} = 1$.
Each of the 9 bosonic matrices $X^I$ are traceless Hermitian matrices and has three components; we can use SO(9) $\times$ SU(2) invariance to write %
\begin{equation} \label{eq:sdef}
X^1=s_1T_1,\quad X^2=s_2T_2,\quad X^3=s_3T_3,\quad X^{4,\dots,9}=0,\qquad
s_1\geq s_2\geq s_3\geq0.
\end{equation}
To use an invariant language,  $s_1, s_2, s_3$ can be viewed as the singular values of the $3\times9$ matrix $X_a^I$.
The wavefunction then depends only on these $s_a$, and
\begin{equation} \label{potential}
V=-\frac14\sum_{I,J}\Tr[X^I,X^J]^2=\sum_{a<b}s_a^2s_b^2,%
\end{equation}
The Jacobian for this change of variables gives $d\mu =\prod_{a=1}^3 s_a^6\prod_{a<b}\big(s_a^2-s_b^2\big)\,d^3s$. Along a valley of the potential $V$,
$s_1\gg s_2,s_3$, we see that $s_2$ and $s_3$ behave like harmonic oscillators with frequency $\sqrt{2} s_1$ to leading order in the Born-Oppenheimer approximation.

In addition to the bosonic variables, we have $16+32=48$ Majorana fermions, yielding a Hilbert space of $\dim \mathcal{H}_f = 2^{24}$. After imposing SO(9) and
SU(2) invariance, the singlet Hilbert space can be represented as
\begin{equation}
\mathcal H_{\rm singlet}\simeq
(L^2(\mathbb R_{+}^3)/S_3) \otimes\mathfrak{F},
\end{equation}
Here $S_3$ permutes the three positive singular values, so the quotient
is represented by $s_1\ge s_2\ge s_3\ge0$ with measure $\mu$, and $\mathfrak{F}$ the finite dimensional subspace
$\mathfrak{F}\subset\mathcal H_f$  described below.

\paragraph{Residual gauge symmetry}
Since we work in a gauge\footnote{In a slight abuse of language, we use ``gauge'' here to refer to both SO(9) and SU(2) invariance.} $X^{4,\dots,9}=0$, there is a residual SO(6) symmetry that preserves $X$. Furthermore, an SO(9) rotation by $\pi$ in the $(a,b)$ plane also leaves $X$ fixed when combined with an SU(2) gauge rotation, where $a,b \in \{1,2,3\}$.\footnote{For example, in the $(1,2)$ plane, with $U=i\sigma_3 \in $ SU(2),
\begin{equation*}
(X^1,X^2)\xrightarrow{\mathrm{SO}(9)}(-X^1,-X^2)
\xrightarrow{\ U(\cdot)U^\dagger\ }(X^1,X^2).
\end{equation*}
} Thus the residual gauge symmetry is 
\begin{align} \label{residualGauge}
H = \mathrm{SO}(6) \times \mathbb{Z}_2 \times \mathbb{Z}_2,
\end{align}
where the last two factors correspond to the independent $\pi$ rotations in the (1,2) and (2,3) planes.
We should demand that the full state is invariant under $H$. Since $H$ acts trivially on the $X$ fields (it is the stabilizer of the bosonic variables $X$), the additional non-trivial constraint comes from its action on the fermionic Hilbert space $\mathcal{H}_f$:
\begin{equation}\label{eq:singlet-stabilizer}
\Sigma_{IJ}\Omega=0\quad(I,J=4,\dots,9),\qquad
 e^{\pi(\Sigma_{ab}+\Lambda_{ab})}\Omega=\Omega\quad(1\leq a<b\leq3).
\end{equation}
Here, the fermionic contributions to the SO(9) and SU(2) generators $C$ in \eqref{eq:QandC} are
\begin{equation}
\Sigma_{IJ}=\frac14\psi_{\alpha c}\gamma^{IJ}_{\alpha\beta}\psi_{\beta c},\qquad
\Lambda_{ab}=\psi_{\alpha a}\psi_{\alpha b}.
\end{equation}
{The states satisfying these conditions form the allowed fermion space $\mathcal F$ at a generic point $s_1 > s_2 > s_3>0$.
SO(6) invariance selects 7898 states from the $2^{24}$-dimensional fermion Fock space;
the additional $\mathbb{Z}_2 \times \mathbb{Z}_2$ constraint reduces this to $\dim\mathcal F=2066$ (see Appendix~\ref{app:fermions}).
Choosing a basis $\ket{t}$ in $\mathcal F$, the singlet wavefunction is specified by 2066 functions,
$\Omega(s)=\sum_{t}f_t(s_1,s_2,s_3)\ket{t}$.}

\subsection{Three-point amplitude and the tail}\label{sec:3pt}
\subsubsection{Tail of the wavefunction}\label{tail}
The tail of $\ket\Omega$ is the region where the two D0-branes are far apart. Consider the following parametrization,
\begin{equation}\label{eq:param}
X^I=\begin{pmatrix}\frac{x^I}{2} & \frac{u^I+\i v^I}{\rt}\\[3pt] \frac{u^I-\i v^I}{\rt} & -\frac{x^I}{2}\end{pmatrix},\qquad
\psi_\alpha=\begin{pmatrix}\frac{\theta_\alpha}{\rt} & \chi_\alpha\\[3pt] \chi^\dagger_\alpha & -\frac{\theta_\alpha}{\rt}\end{pmatrix}.
\end{equation}
The diagonal entries are the two D0-branes, at $\pm\vec x/2$, with sixteen real fermions $\theta$. The off-diagonal
entries are the open strings between them, with coordinates $u,v\in\R^9$ and sixteen complex fermions $\chi$. 
Unlike the previous section, in this subsection we use the background field gauge~\cite{Becker:1997xw}. In this gauge the moduli space Lagrangian becomes~\cite{Douglas:1996yp},
\begin{equation} \label{eq:Lfamous}
L=\frac{\dot{\vec x}^2}{4}+\frac{15}{16}\frac{\dot{\vec x}^4}{x^7}+\dots
\end{equation}
With $\vec p=\dot{\vec x}/2+\dots$, the Hamiltonian is $
H=p^2-\frac{15\,p^4}{x^7}+\dots
$
At leading order the two D0-branes are governed by the free non-relativistic Hamiltonian
$H_{0}=-\nabla_{\vec x}^2 .$
Writing $R_\ell=x^{-4}f_\ell$ gives the radial equation
\begin{equation}\label{eq:radial}
\Bigg[\!-\frac{d^2}{dx^2}+\frac{(\ell+3)(\ell+4)}{x^2}\Bigg]f_\ell=E\,f_\ell .
\end{equation}
The second term is the angular momentum barrier.
The wavefunction then becomes $\ket\Omega\simeq\Psi(\vec x)\ket{0_{\hat x}}$,
 (normalized so that $|\Psi|^2d^9x$ is
the probability, counting both $\vec x$ and $-\vec x$). Under SO(9) the $256$ polarization states split as
\begin{equation}
256=\mathbf{44}\oplus\mathbf{84}\oplus\mathbf{128},
\end{equation}
the symmetric traceless tensor $\mathbf{44}$ ``graviton'' $\ket{ij}$, the three-form and the ``gravitino''. The total angular momentum
of the bound state splits into two terms,
\begin{equation}
J=L_{\rm rel}+S_{\rm rel},
\end{equation}
 $L_{\rm rel}$ is the orbital angular momentum of $\vec x$ and $S_{\rm rel}$ rotates $\theta$. 
 Recall that $\ket\Omega$ is an SO(9) singlet $J=0$.
 Since the 256 polarization states contain no singlet, we need non-zero
orbital angular momentum to combine with the spin to satisfy 
$(L_{\rm rel}+S_{\rm rel})\Psi=0$. Orbital angular momentum always comes in symmetric traceless tensor
representations, so we have to choose $\ell=2$ or the $\mathbf{44}$,
\begin{equation}
\Psi(\vec x)
=\frac{f(x)}{x^4}\,\hat x_i\hat x_j\ket{ij}.
\end{equation}
Here $\Psi(\vec x) = \braket{x}{\Omega}$ is the position-space wavefunction, valued in the 256 dimensional Hilbert space associated with the fermion zero modes.
At $E=0$ and $\ell=2$ the radial equation~\eqref{eq:radial} gives  $R_{\ell} =x^{-4}f\propto x^{-9}$. We have then the leading order part of the ground state wavefunction as,
\begin{equation}\label{eq:tail}
\Psi(\vec x)\simeq g\,\pd_i\pd_j\mathcal G_9(x)\ket{ij}=\frac{945\,g}{32\pi^4}\,\frac{\hat x_i\hat x_j}{x^9}\ket{ij},\qquad
\mathcal G_9(x)=\frac{15}{32\pi^4x^7}.
\end{equation}
The complete wavefunction at large separation then is
\begin{equation}
\frac{945\,g}{32\pi^4}\,\frac{\hat x_i\hat x_j}{x^9}\ket{ij}\otimes\ket{0_{\hat x}},
\end{equation}
and the probability per unit $x$ is
\begin{equation}
P(x)=\frac{945\,g^2}{4\pi^4 x^{10}}
\end{equation}
To go beyond leading order we include the $p^4/x^7$ term of $H$. Acting on the leading solution $\Psi_0\propto\hat x_i\hat x_jx^{-9}\ket{ij}$ of \eqref{eq:tail}, each derivative gives a factor $1/x$, so,
\begin{equation}
    \frac{p^4}{x^7}\Psi_0\sim x^{-11}\Psi_0
\end{equation}
 Solving $H\Psi=0$ to first order in this term gives $\Psi=\Psi_0\big(1+O(x^{-9})\big)$, so the probability per unit $x$ is still
\begin{equation}
P(x)\,dx=\frac{A}{x^{10}}\big(1+O(x^{-9})\big)\,dx,\qquad A=\frac{945\,g^2}{4\pi^4}.
\end{equation}
In the variables of the previous section, the radial coordinate is $r=\rt\,s_1$, with $s_1$ the largest singular value of the $3 \times 9$ matrix $X_a^I$. This is the largest distance between the two eigenvalues of $\hat n^IX^I$ over unit vectors $\hat n$. For $\hat n=\hat x$, \eqref{eq:param} gives
\begin{equation}
\hat x^IX^I=\frac{x}{2}\,\sigma_3+\frac{\hat x\cdot u}{\rt}\,\sigma_1-\frac{\hat x\cdot v}{\rt}\,\sigma_2,
\end{equation}
and the distance between its two eigenvalues gives
\begin{equation}
r^2=x^2+2(\hat x\cdot u)^2+2(\hat x\cdot v)^2+\dots
\end{equation}
In this gauge $u$ and $v$ also oscillate along $\vec x$ with frequency $x$, so in the ground state
\begin{equation}
\langle(\hat x\cdot u)^2\rangle=\langle(\hat x\cdot v)^2\rangle=\frac{1}{2x}.
\end{equation}
So $r=x+1/x^2+\dots$, and changing variables in $P(x)\,dx$,
\begin{equation}
P(r)=A\Big(r-\frac{1}{r^2}\Big)^{-10}\Big(1+\frac{2}{r^3}\Big)+\dots=\frac{A}{r^{10}}\Big(1+\frac{12}{r^3}+\dots\Big).
\end{equation}
We can continue the expansion to the next order, yielding (see Appendix \ref{r_6_tail}) %
\begin{equation}\label{subleading_tail}
P(r)=\frac{A}{r^{10}}\Big(1+\frac{12}{r^3}+\frac{435}{r^6}+O(r^{-9})\Big).
\end{equation}
While this fixes the relative coefficients of the subleading terms, it does not fix the overall coefficient $A$. In the next subsection we relate $A$ to the protected three-point amplitude computed in~\cite{Herderschee:2023pza}. In Section~\ref{sec:checks} we compute $A$ and the subleading coefficients from the numerical wavefunction and compare.
\subsubsection{Relation to the three-point amplitude}\label{sec:3pt-LS}
We now relate the coefficient $g$ in~\eqref{eq:tail} to the amplitude for two D0-branes to join into the bound state. 
We use the coordinates in \eqref{eq:param} with $\vec x$ replaced by $\vec r$ and impose the transverse condition
\begin{equation}\label{eq:gauge3pt}
\hat r\cdot u=\hat r\cdot v=0.
\end{equation}
Here $r=\sqrt2\,s_1$. At leading order, the sixteen
transverse components of $u,v$ and the fermions $\chi$
are oscillators of frequency $r$, with ground state
$\ket{0_{\hat r}}$.

\paragraph{Factorization of the 4-pt amplitude.}
Let $\ket{\vec p,a}$ denote the free two-D0-brane state with wavefunction $e^{\i\vec p\cdot\vec r}\ket{a}\otimes\ket{0_{\hat r}}$, where $a$ labels the relative fermion polarization\footnote{Technically, exchanging the two D0-branes is a residual gauge transformation. It sends $\vec r\to-\vec r$ and $\theta\to-\theta$, with the latter acting on the polarizations as fermion parity $\Pi$. The normalized gauge-invariant states are
\begin{equation}\label{eq:exchange}
\ket{\vec p,a}_{\rm sym}=\frac{\ket{\vec p,a}+\Pi\ket{-\vec p,a}}{\rt},\qquad \Pi=+1\ \text{on}\ \mathbf{44}\oplus\mathbf{84},\qquad \Pi=-1\ \text{on}\ \mathbf{128}.
\end{equation}
Let $\pi$ be the projector onto the physical (gauge-invariant) Hilbert space. Then $\pi \ket\Omega = \ket\Omega$, so $\mathcal{M}(\vec{p},a) = \bra\Omega H_\text{int} \pi \ket{\vec{p},a} = \bra\Omega H_\text{int} \ket{\vec{p},a}_\text{sym} / \sqrt{2}$.
}. We write $H=H_0+H_{\rm int}$, with $H_0=-\nabla_{\vec r}^2$ and $H_0\ket{\vec p,a}=p^2\ket{\vec p,a}$. Here $H_{\rm int}$ includes the interactions and the dependence of the oscillator ground state on $\vec r$. The ground-state equation $H\ket\Omega=0$ gives the Lippmann--Schwinger equation for the bound state
\begin{equation}\label{eq:LS}
\braket{\Omega}{\vec p,a}=-\frac{\mathcal M(\vec p,a)}{p^2},\qquad\mathcal M(\vec p,a)\equiv\bra\Omega H_{\rm int}\ket{\vec p,a}.
\end{equation}
To see how this overlap $\mathcal{M}$ enters scattering, consider the transition operator $T(E)=H_{\rm int}+H_{\rm int}(E-H+\i 0)^{-1}H_{\rm int}$. 
Its matrix elements are related to the $2\to 2$ $S$-matrix by
\begin{equation}
\bra f S\ket i
=\braket{f}{i}
-2\pi\mathrm i\,\delta(E_f-E_i)
\bra f T(E_i)\ket i .
\end{equation}
Since the bound state has zero energy, the resolvent
contains $\ket\Omega\bra\Omega/(E+\mathrm i0)$.
Its contribution to the scattering amplitude is therefore
\begin{equation}
\bra{\vec p\,',a'}T(E)\ket{\vec p,a}
\supset
\frac{\mathcal M(\vec p\,',a')^*
      \mathcal M(\vec p,a)}
     {E+\mathrm i0}.
\end{equation}
The two factors in the numerator describe the incoming pair joining into the bound state and the bound state splitting into the outgoing pair. We should evaluate near the on-shell condition $\vec{p}^2=0$. In other words, this equation is just saying that the 4-pt amplitude factorizes into a product of 3-pt amplitudes. We therefore identify
\begin{align}\label{eq:Agq}
    \mathcal A_\text{3pt} =\bra\Omega H_{\rm int}\ket{\vec p,a}\big|_{p^2=0}=\mathcal M(\vec p,a)\big|_{p^2=0}.
\end{align}
Note that we are doing a scattering computation in quantum mechanics and are not assuming any relativistic symmetry of the amplitude.

\paragraph{Relating $\M$ to $g$.}
We now expand $\mathcal M$ at small $\vec{p}^2$ and relate it to the normalization of the power law. 
The overlap in~\eqref{eq:LS} is the Fourier transform of the component of $\ket\Omega$ with the oscillators in their ground state,
\begin{equation}\label{eq:gammaint}
\mathcal M(\vec p,a)=-p^2\int d^9r\,e^{\i\vec p\cdot\vec r}\braket{\Psi(\vec r)}{a},\qquad\Psi(\vec r)=\braket{0_{\hat r}}{\Omega}.
\end{equation}
Its leading tail is
\begin{equation}\label{eq:tailr}
\Psi(\vec r)\simeq \frac{945\,g}{32\pi^4}\,\frac{\hat r_i\hat r_j}{r^9}\ket{ij}.
\end{equation}
To find the form of the corrections, we use the large-$r$ expansion of the full wavefunction~\cite{Lin:2014wka},
\begin{equation}\label{eq:Omegaexp}
\ket\Omega=\ket{\Omega_0}+\ket{\Omega_1}+\ket{\Omega_2}+\dots,\qquad \ket{\Omega_n}\sim r^{-3n/2}\ket{\Omega_0},
\end{equation}
where $\ket{\Omega_0}$ is~\eqref{eq:tailr} times $\ket{0_{\hat r}}$. Rescaling $(u,v)=r^{-1/2}(\tilde u,\tilde v)$ gives
\begin{equation}
H=r\,H_{\rm osc}+r^{-1/2}H_{\rm Y}+\dots.
\end{equation}
Here $H_{\rm osc}$ is the oscillator Hamiltonian and $H_{\rm Y}$ is the Yukawa interaction, linear in $(\tilde u,\tilde v)$, in $\chi$ and in $\theta$. It excites one bosonic and one fermionic oscillator. Hence the first correction,
\begin{equation}\label{eq:Omega1}
\ket{\Omega_1}=-\frac{1}{r^{3/2}}\,H_{\rm osc}^{-1}H_{\rm Y}\ket{\Omega_0},
\end{equation}
has zero overlap with $\ket{0_{\hat r}}$. The inverse here acts on the excited oscillator states.

The same conclusion holds at every odd order. Since $H$ is polynomial in $(u,v)$ with integer powers of $r$, half-integer powers of $r$ occur with odd powers of $(\tilde u,\tilde v)$. Their projection onto the even oscillator ground state vanishes. At odd $n$, a component along $\ket{0_{\hat r}}$ would therefore have to obey the free equation,
\begin{equation}\label{eq:Psin}
H_0\Psi_n=0,\qquad \Psi_n(\vec r)=\braket{0_{\hat r}}{\Omega_n}.
\end{equation}
The only decaying solution with $\ell=2$ falls as $r^{-9}$, already the order of $\Psi_0$. It cannot appear at an odd subleading order, so $\Psi_n=0$ for odd $n$.

Thus the projected tail has relative corrections in powers of $r^{-3}$. Denote the first two coefficients by $\xi_3$ and $\xi_6$. Components with excited oscillators do not enter~\eqref{eq:gammaint}, but their norm contributes to $P(r)$ in~\eqref{subleading_tail}. For example, $\braket{\Omega_1}{\Omega_1}\sim r^{-3}\braket{\Omega_0}{\Omega_0}$. The contribution from large $r$ to~\eqref{eq:gammaint} is
\begin{align}
\mathcal M(\vec p,a)&=-p^2\,\frac{945\,g}{32\pi^4}\int d^9r\,e^{\i\vec p\cdot\vec r}\,\frac{\hat r_i\hat r_j}{r^9}\left(1+\frac{\xi_3}{r^3}+\frac{\xi_6}{r^6}+\dots\right)\braket{ij}{a}\\
&=g\,p_ip_j\braket{ij}{a}\left(1-\frac{21\pi}{2048}\,\xi_3\,p^{3}-\frac{\xi_6}{34320}\,p^6\log p^2+\dots\right).
\end{align}
Writing
\begin{equation}\label{eq:gammaform}
\mathcal M(\vec p,a)=\kappa(p^2)\,p_ip_j\braket{ij}{a},
\end{equation}
we see that $\kappa$ is finite at $p^2=0$. The  corrections vanish as $(p^2)^{3/2}$ and $(p^2)^3\log p^2$. The region of finite $r$ adds only terms analytic in $p^2$, starting at order $p^2$, so these also vanish there. Thus $\kappa(0)=g$, and only the leading tail contributes,
\begin{equation}\label{eq:vertex}
\mathcal M(\vec p,a)\big|_{p^2=0}=g\,p_ip_j\braket{ij}{a}.
\end{equation}

\paragraph{Complex kinematics and polarization}
For real momenta, $p^2=0$ forces $\vec p=0$. To obtain a nonzero amplitude we continue to complex momenta,
\begin{equation}
\vec p=q\epsilon,\qquad \epsilon=\frac{e_9+\i e_8}{\rt},\qquad \epsilon^2=0,\qquad
\ket-=\epsilon_i^*\epsilon_j^*\ket{ij}.
\end{equation}
Then using \eqref{eq:Agq},
\begin{equation}\label{eq:Agq2}
\mathcal A_\text{3pt}=\mathcal M(\vec p,a)\big|_{p^2=0} = g\,q^2\braket{-}{a}.
\end{equation}
The amplitude computed in~\cite{Herderschee:2023pza}, converted to our units and state normalization in Appendix~\ref{app:HM}, gives
\begin{equation}\label{eq:g3pt}
g_{\rm 3pt}=16\pi^2,\qquad A_{\rm 3pt}=\frac{945\,g_{\rm 3pt}^2}{4\pi^4}=60480.
\end{equation}
We compare this with the tail of the numerical wavefunction in Section~\ref{sec:checks}.

\subsection{Numerical solution}\label{ground_solve}

To determine the ground state $\ket\Omega$, we must solve for 2066 functions $f_t(s_1,s_2,s_3)$ of three variables. 
We will solve the ground state wavefunction by using a variant of the variational method. Let us introduce the variational functional $\mathcal{F}/\mathcal{N}$,
\begin{equation}\label{eq:ground-least-squares}
\mathcal{F}[\Omega]=\sum_{\alpha=1}^{16}\bra{\Omega} Q_\alpha w(s) Q_\alpha \ket\Omega%
\qquad
N[\Omega]= \braket{\Omega}{\Omega}%
\end{equation}
The naive variational method  would correspond to minimizing $H$ or setting $w(s) = 1$.
However, since the ground state is supposed to be annihilated by all the supercharges, the variational functional should vanish if we find the true ground state for any $w(s)$.
We will use this to our advantage; in particular, we will choose a $w(s)$ that gives good convergence of the tails. In particular, since the radial probability density decays at large $r$ as $P(r) \sim r^{-10}$ \cite{Polchinski:1999br, Lin:2014wka}, a large relative error in the tail can contribute little to the unweighted $\mathcal{F}$. A related difficulty is that we can form a wavepacket out of scattering states that is localized to a region $r\sim R$ and $\Delta p\sim R^{-1}$. For $w=1$ this gives $\mathcal{F}/\mathcal{N}\sim R^{-2}$. Thus for $w=1$ we can be contaminated by such low energy scattering states.

Motivated by these considerations, we chose
\begin{equation}\label{eq:ws}
w(s)=\left(1+\frac{s_1^2}{c_w}\right)^2,\qquad c_w=14.29.
\end{equation}
Since $r=\sqrt2\,s_1$, this gives $w\propto r^4$ at large separation. The tail then receives a larger weight, while these low energy wavepackets would give $\mathcal{F}/\mathcal{N}\propto R^4R^{-2}=R^2$ at large $R$. The scale $c_w$ sets where the extra weighting starts.\footnote{This is chosen by a rough estimate of where the core transitions to tail.}

 From \eqref{eq:sdef}, note that $\Tr X^2= s^2 = s_1^2 + s_2^2 + s_3^2 $. We choose the ansatz \begin{equation}\label{eq:trial}
\begin{aligned}
\Omega(s)&=\frac{1}{(1+s^2/c)^{3}}\exp\left({-\frac{V(s)}{\sqrt{2 s^2}}} \right)\sum_{t}\chi_{t}(s)\,\ket{t},%
\end{aligned}
\end{equation}
Here $\Omega$ is a 2066-dimensional vector which can be expanded in the $\ket{t}$ basis. We describe  $\chi_t(s)$ in Appendix \ref{app:ground-numerics}; {we use $808\,960$ points in the three coordinates $(s_1,s_2,s_3)$, with 2066 fermionic components at each point, giving $\sim 1.7\times10^9$ unknowns in total.} The variational parameters are precisely just the values of $\chi_t$ at these discrete locations; away from these locations, $\chi_t$ is obtained by polynomial interpolation \eqref{eq:polynomialInterpolation} in some suitable coordinate system.

This ansatz was chosen to have the following property: along a valley, the potential $V$ in \eqref{potential} gives an exponential factor $\exp( -s_1(s_2^2+s_3^2)/\rt)$ , which is precisely the Gaussian ground state of the off-diagonal oscillators to leading order. We chose $c = 20/2^{4/3}$.  %
Since both factors are non-zero at finite $s$, passing from $f_t$ to $\chi_t$ is just a change of the unknown functions. We do not require $\chi_t$ to approach a constant, and the tail amplitude is determined by the solution.

Let $\mathbf c$ be the vector of values of $\chi_t$ at all nodes and
for all components $t$, and let $A$ and $M$ be the matrices representing
the functional $\mathcal{F}=\mathbf c^\dagger A\mathbf c$ and the norm
$N=\mathbf c^\dagger M\mathbf c$, respectively. The coefficients that
minimize $\mathcal{F}/\mathcal{N}$ are then given by the eigenvector of
$A\mathbf c=\lambda M\mathbf c$ with the smallest eigenvalue. This allows us to solve the variational problem efficiently by solving a generalized eigenvalue problem.

To solve this generalized eigenvalue problem, we start with the 200 linear combinations of fermion states that carry the most norm in the current wavefunction. We add combinations obtained from the derivatives of $\mathcal{F}/\mathcal{N}$ with respect to the remaining components, using up to 230 combinations in total. We solve the eigenvalue problem in this smaller space, then choose the combinations again as the wavefunction improves. The selection itself uses all 2066 components.

We then hold the core fixed and minimize ${\mathcal F}$ over the outer grid, allowing all 2066 components to vary.  The core enters through the term comparing the two polynomials at $s_1=5.67$, so it fixes the amplitude of the tail relative to the core. These updates improve the tail while leaving the core unchanged, apart from the overall normalization. Details of the iteration are given in App.~\ref{app:ground-numerics}.

The sum of squares also avoids subtracting large energy contributions. Along a valley, each of the sixteen bosonic oscillators contributes $r/4$ to both the kinetic energy $K$ and the potential $V$, while the fermion energy is $-8r$. Thus
$
\evs{K}+\evs{V}+\evs{F}\simeq4r+4r-8r=0.$
A fractional underestimate $\epsilon>0$ of $K$ would give a false energy $-4\epsilon r$, which could be lowered by moving the state to larger $r$. %

At a boundary $s_a=s_b>0$, smoothness requires $(\Sigma_{ab}+\Lambda_{ab})\Omega=0$. We retain the singular angular terms in the supercharge equations, so a violation is penalized by $\mathcal{F}$ at points close to the wall. We also put a small lower bound on the numerical weights in $\mathcal{F}$ so that points with very small measure are not ignored. We inspect the equation errors near the transverse edges and repeat the solve on finer grids. This controls the condition numerically. App.~\ref{app:ground-numerics} gives the implementation and the checks.

\subsection{Checks and results}\label{sec:checks}
\begin{figure}[htbp]
\centering
\includegraphics[width=\textwidth]{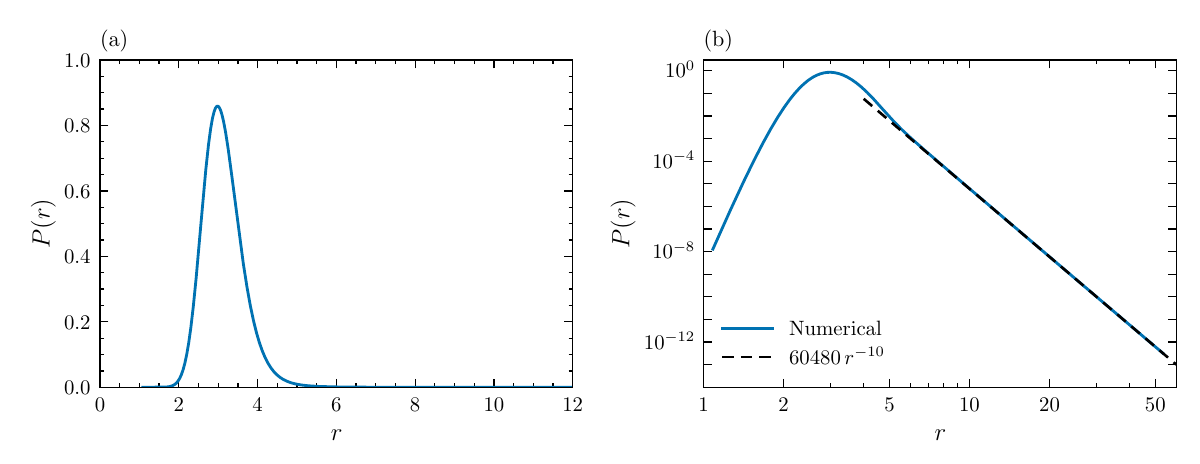}
\caption{Radial probability density, with $r=\sqrt2\,s_1$ the separation between the two D0-branes. (a) Linear scale near the maximum. (b) Logarithmic scales over the numerical range. The dashed curve is the leading tail $A_{\rm 3pt}r^{-10}$ of~\eqref{eq:g3pt}, from the three-point amplitude~\cite{Herderschee:2023pza}, shown for comparison. The numerical curve is shown for $1.08\le r\le53.4$.}
\label{fig:radial}
\end{figure}

\begin{figure}[htbp]
\centering
\includegraphics[width=\textwidth]{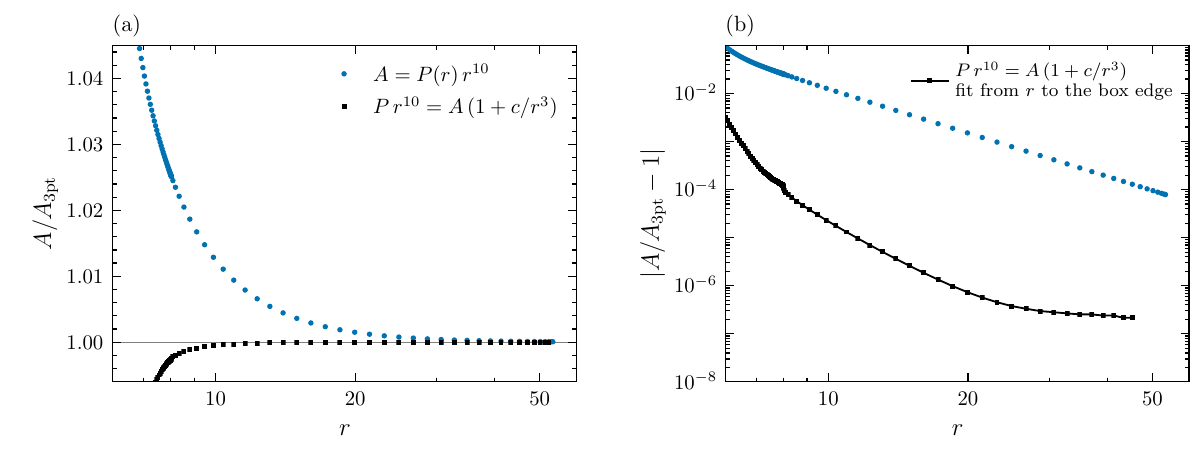}
\caption{Extracting the three-point amplitude from the power law tail. (a) $A/A_{\rm 3pt}$, with $A_{\rm 3pt}=60480$. Blue circles use $A=P(r)r^{10}$ at each radial grid point. Black squares fit $P(r)r^{10}=A(1+c/r^3)$ on five neighboring radial grid points, leaving both $A$ and $c$ free.  (b) $|A/A_{\rm 3pt}-1|$ for $A=P(r)r^{10}$ (blue) and for one fit of $P(r)r^{10}=A(1+c/r^3)$ over every radial grid point from $r$ out to the edge of the box (black). Beyond $r=25$ that fit stays within $3.4\times10^{-7}$ of $A_{\rm 3pt}$.}
\label{fig:tail-amplitude}
\end{figure}

Since our variational procedure used a non-trivial weight $w$, a first cross-check is just to evaluate the energy\footnote{Since the numerics is effectively done in a finite but very large box and with a finite resolution, supersymmetry is slightly broken. Therefore, there is a slight difference numerically between evaluating $E_Q$ and $\ev{H}$. Evaluating the Hamiltonian 
$\evs{H} = \evs{K}+\evs{V}+\evs{F}=1.6\times10^{-9}$, differing from
$E_Q$ by $2.3\times10^{-9}$.}  %
$
E_Q\equiv \frac{1}{16N}\sum_{\alpha=1}^{16} \bra{Q_\alpha \Omega}\ket{Q_\alpha\Omega} %
\simeq3.9\times10^{-9}.$
We also check the kinetic, potential and fermion energies of the numerical state. The virial theorem gives $2\evs{K}=4\evs{V}+\evs{F}$ for a normalizable energy eigenstate. Together with $\evs{H}=0$, this implies
\begin{equation}
\evs{K}=\evs{V}=-\tfrac12\evs{F}.
\end{equation}
For the numerical ground state, we find agreement with the virial theorem to $3\times10^{-8}$ accuracy \footnote{The solve on a coarser grid, with 116 radial and $26\times22$ transverse grid points, gives every moment in Table~\ref{tab:moments} unchanged at the precision shown.},
 see Table~\ref{tab:energies}. %
\begin{table}[tbp]
\centering
\begin{tabular}{cccc}
\toprule
$\evs{K}/g_{\rm YM}^{2/3}$ & $\evs{V}/g_{\rm YM}^{2/3}$ & $\evs{F}/g_{\rm YM}^{2/3}$ & $E /g_{\rm YM}^{2/3}$\\
\midrule
$17.1124633$ & $17.1124638$ & $-34.2249271$ & $1.6\times10^{-9}$\\
\bottomrule
\end{tabular}
\caption{Kinetic, potential and fermionic energies of the numerical ground state and its total energy.}
\label{tab:energies}
\end{table}
We next examine the wavefunction at large D0-brane separation $r=\sqrt2\,s_1$. Integrating $\Omega^\dagger\Omega$ over the remaining variables gives the radial probability density
\begin{equation*}
P(r)=\int d\mu\,\Omega^\dagger\Omega\,\delta\big(r-\sqrt2\,s_1\big).
\end{equation*}
Thus $P(r)\,dr$ is the probability of a separation between $r$ and $r+dr$. Figure~\ref{fig:radial} shows the density near its maximum and over the full numerical range.
The analytical expectation for the leading relative wavefunction is $r^{-9}$~\cite{Lin:2014wka}. Following~\eqref{subleading_tail},
\begin{equation}
P(r) \approx\frac{A}{r^{10}}\left(1+\frac{12}{r^3}+\frac{435}{r^6}+\frac{23\,092}{r^9}+\cdots\right),
\end{equation}
 The numerical density  we find agrees very well  with this expansion.\footnote{With all three corrections included and only the overall normalization fitted, the largest relative difference is $3.4\times10^{-6}$ over $10\le r\le53.5$, and $4.7\times10^{-7}$ over $20\le r\le53.5$ with the same normalization.}
We determine the powers from the numerical density by fitting
\begin{equation*}
P(r) \approx A\,r^{-\tau}\big(1+c_1r^{-\tau_1}+\cdots\big).
\end{equation*}
Allowing the powers to vary, the exponent scan gives $\tau=10 \pm 2 \times 10^{-5}$ and $\tau_1=3.00\pm0.02$. In particular, the three-term fit $P(r)=A r^{-\tau}(1+c_1r^{-\tau_1}+c_2r^{-\tau_2})$ on $10.69\le r\le53.4$, out to the edge of the box, with all exponents and coefficients free, gives $\tau=10\pm2\times10^{-5}$ and $\frac{A}{A_{\rm 3pt}}-1=-1.3\times10^{-7}.$
 With the powers fixed, we fit $P(r)r^{10}=A(1+c_1/r^3+c_2/r^6)$ on windows starting at $r\simeq11,13,18$ and ending at $53.5$. The coefficients agree with 12 and 435 with relative errors below $8\times10^{-4}$ and $9\%$, respectively, and the fitted $A$ gives the three-point coupling\footnote{The quoted error holds these powers fixed. It includes changes of grid, box size, fit window and iteration, and a tail-truncation error estimated by applying the same fits to the analytic tail.},
\begin{equation}
    g=157.9136\pm0.0003, \qquad g_{\rm{3pt}} = 16\pi^2\ (157.913670)
\end{equation}
 Figure~\ref{fig:tail-amplitude} shows the amplitude extracted locally from the numerical density.

\begin{figure}[htbp]
\centering
\includegraphics[width=0.85\textwidth]{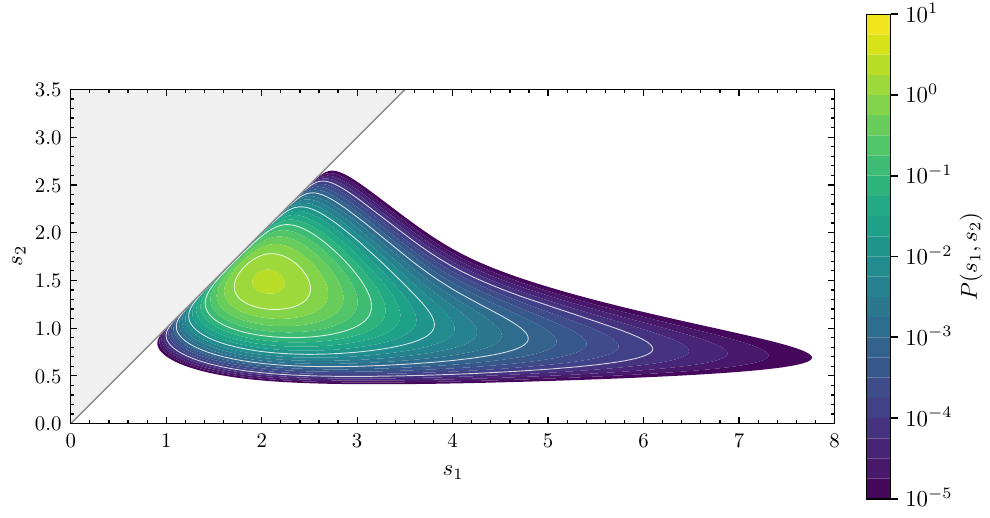}
\caption{Joint ground-state probability density $P(s_1,s_2)$ for the two
largest singular values, obtained by integrating $|\Omega|^2$ over $s_3$ and summing over fermionic indices. The color scale is logarithmic. Note that $s_2$ is a measure of how much the matrices fail to commute \eqref{potential}. The singular values are ordered $s_1\geq s_2$ so the gray region is unphysical. At large $s_1$,
the density narrows towards small $s_2$ (the commuting valley of the potential $V$). The off-diagonal modes
have frequency $\sqrt{2}\,s_1$, so their width decreases
$\sim s^{-1/2}$ as they are frozen into their ground state.}
\label{fig:gs-s1-s2}
\end{figure}

We next compute expectation values in the numerical state. Higher moments are particularly sensitive to the tail. Along a valley, $\Tr X^{2n}\propto r^{2n}$ and $P(r)\propto r^{-10}$, so the contribution from separations beyond $R$ scales as
\begin{equation*}
\int_R^\infty dr\,P(r)\,r^{2n}\propto R^{2n-9},\qquad n=1,\ldots,4.
\end{equation*}
In particular, the omitted contributions to $\evs{\Tr X^6}$ and $\evs{\Tr X^8}$ decrease only as $R^{-3}$ and $R^{-1}$. We integrate the numerical wavefunction up to $R=53.5$ and add the contribution beyond it from a fit to the tail, which supplies 1.1\% of $\evs{\Tr X^8}$. Table~\ref{tab:moments} compares the resulting expectation values with independent level-14 bootstrap bounds. All nine values lie within the allowed intervals.

\begin{table}[htbp]
\centering
\begin{tabular}{lcc}
\toprule
 & wavefunction & bootstrap bound \\
\midrule
$\evs{\Tr X^2}$ & 8.074210 & $[8.074159,\ 8.074275]$ \\
$\evs{\Tr X^8}$ & 800.617 & $[800.573,\ 800.628]$ \\
$\evs{V}$ & 17.11246 & $[17.112328,\ 17.112636]$ \\
$\evs{V\,\Tr X^2}$ & 149.5830 & $[149.58214,\ 149.58344]$ \\
$\evs{V^2}$ & 344.790 & $[344.78618,\ 344.79384]$ \\
$\evs{F\,\Tr X^2}$ & $-283.5589$ & $[-283.55970,\ -283.55739]$ \\
$\evs{F\,V}$ & $-612.7787$ & $[-612.78653,\ -612.77247]$ \\
$\evs{\det M}$ & 9.979125 & $[9.9790178,\ 9.9791790]$ \\
$\evs{P^I_a\,\Tr X^2\,P^I_a}$ & 294.9519 & $[294.95051,\ 294.95254]$ \\
\bottomrule
\end{tabular}
\caption{Moments of the numerical ground state and level 14 bootstrap bounds on them. Here we use the shorthand $X^{2n}\equiv(X^IX^I)^n$ and $M_{ab}=X^I_aX^I_b$. $F$ and $V$ are the fermionic and bosonic potential terms in the Hamiltonian.
}
\label{tab:moments}
\end{table}

\section{S-matrix}\label{sec:smatrix}

\subsection{SUSY multiplet \label{sec:supermultiplet}}
 We first describe the supersymmetry multiplet of a state with energy $E>0$. 
We group the 16 real supercharges $Q_\alpha$ into eight complex supercharges
\begin{equation}
a_j=\frac{Q_{2j-1}+\i Q_{2j}}{2\sqrt E},\qquad
a_j^\dagger=\frac{Q_{2j-1}-\i Q_{2j}}{2\sqrt E},\qquad
\{a_i,a_j^\dagger\}=\delta_{ij},\qquad j=1,\ldots,8.
\end{equation}
Starting from a state $\ket{E,r;0}$ annihilated by every $a_j$, acting with the $a_j^\dagger$ generates $2^8=256$ states at the same energy. Here the top state transforms in some irrep $R$ of SO(9) (sometimes referred to as superspin) and $r$ is a label that takes $\dim R$ values. %
Hence the complete energy eigenspace can be written
\begin{equation} \label{eq:supermultiplet}
\mathcal H_E=S\otimes  R,\qquad
S=\mathbf{44}\oplus\mathbf{84}\oplus\mathbf{128}.
\end{equation}
Here $S$ is the 256-dimensional space generated by the supercharges\footnote{The $\mathbf{84}$ is the representation of completely antisymmetric three-index tensors,
\begin{equation}
A_{ijk}=A_{[ijk]},\qquad i,j,k=1,\ldots,9,\qquad
\dim\mathbf{84}=\binom{9}{3}=84.
\end{equation} while $\mathbf{128}$ is the fermionic ``gravitino'' irrep.}. 
So an irreducible representation (a supermultiplet) of the SUSY algebra is labeled by $S\otimes R$, containing $256\dim R$ states. This analysis of the irreps applies to both scattering states and bound states with $E>0$.

\subsection{Scattering kinematics}

We  apply the methods  of the previous section to the $2 \to 2$ scattering.
In the SU(2) theory, the scattering takes place in the center-of-mass frame:
\begin{align}
\vcenter{\hbox{\begin{tikzpicture}[>=stealth, thick]
  \draw[-<] (0,0) -- (2.5,0)
    node[right] {$- \mathbf{p}$ (in)};
  \draw[-<] (0,0) -- (-2.5,0)
    node[left] {$ \mathbf{p}$ (in)};
  \draw[->] (0,0) -- (40:2.5)
    node[above right] {$ \mathbf{p}'$ (out)};
  \draw[->] (0,0) -- (220:2.5)
    node[below left] {$ -\mathbf{p}'$ (out)};
  \draw (0.7,0) arc[start angle=0,end angle=40,radius=0.7];
  \node at (20:0.95) {$\theta$};
\end{tikzpicture}}}
\end{align}
The two invariants are $ E=p^2$, and the scattering angle depicted above. Equivalently, we can use the relativistic notation (suitable for connecting with the BFSS conjecture) $s = 4 E =4p^2,$ and $t=-2p^2(1-\cos\theta)$.
The asymptotic states of the SU(2) theory are labelled by a single momentum in the center-of-mass frame, together with a discrete label
\begin{equation}
    |i\rangle  =|\mathbf{p},a\rangle, \qquad |f\rangle = |\mathbf{p}',b\rangle
\end{equation}
where $a, b$ label states in the polarization space $S$ of the relative fermionic zero modes, defined in \eqref{eq:supermultiplet}. The S-matrix in this basis is,
\begin{equation}
    S_{ba}(\bold{p}',\bold{p}) = \langle \bold{p}',b; \text{out}| \bold{p},a;\text{in}\rangle
\end{equation}
  
Equivalently, we may think of the $S$ matrix as a function of the center-of-mass energy $E= p^2$ and the two directions $\mathbf{n}, \mathbf{n}'$. %

Equivalently, at fixed energy $E=p^2$, we may label
the asymptotic states by $|E,\mathbf n,a\rangle$, where
$\bold p=\sqrt E\,\mathbf n$ and $\mathbf n\in S^8$.
The momentum-space wavefunction can therefore be written
as $\psi_a(E,\mathbf n)$; at fixed $E$ we abbreviate it
to $\psi_a(\mathbf n)$.\footnote{The change of variables
has measure
$d^9p=\tfrac12 E^{7/2}\,dE\,d\Omega_{\mathbf n}$.
We absorb the corresponding normalization factor into
the energy-angle states and their wavefunctions, so that
the angular inner product uses $d\Omega_{\mathbf n}$.
The fixed-energy S-matrix below has the
energy-conserving delta function factored out.}
The S-matrix maps the incoming angular wavefunction to
the outgoing one:
\begin{equation}
\psi_b^{\mathrm{out}}(\mathbf n')
=
\int_{S^8}d\Omega_{\mathbf n}\,
S_{ba}(E;\mathbf n',\mathbf n)
\psi_a^{\mathrm{in}}(\mathbf n).
\end{equation}
\paragraph{1. Remove the polarization dependence.}
Supersymmetry relates scattering amplitudes with different
polarizations; %
conservation of the asymptotic supercharges
requires
\begin{equation} \label{ward_susy}
Q_\alpha^0(\mathbf p')S 
=
S Q_\alpha^0(\mathbf{p}).
\end{equation}
Let $W(\mathbf n)$ be the unitary operator satisfying
\begin{equation}\label{eq:smatrix-W}
W(\mathbf n)\theta_\alpha W(\mathbf n)^{-1}
=
(\mathbf n\cdot\gamma)_{\alpha\beta}\theta_\beta,
\qquad
W(\mathbf n)^2=1.
\end{equation}
The solution to the Ward identity \eqref{ward_susy} is 
\begin{equation}\label{eq:smatrix-factorization}
S_{ba}(E;\mathbf n',\mathbf n)
=
f(E,\cos \theta)
W(\mathbf n')W(\mathbf n)
\end{equation}
The polarization dependence is fixed, while the scalar $f$ contains the dynamics. 
It is convenient to introduce
 \begin{equation}
 \widetilde\psi(\mathbf n)=W(\mathbf n)\psi(\mathbf n),
 \end{equation}
The in/out wavefunctions are then related by
\begin{equation}
\widetilde\psi_a^{\mathrm{out}}(\mathbf n')
=
\int_{S^8}d\Omega_{\mathbf n}\,
f(E,\mathbf n'\cdot\mathbf n)
\widetilde\psi_a^{\mathrm{in}}(\mathbf n).
\end{equation}

\paragraph{2. Decompose into partial waves and use elasticity.}
We now expand the transformed angular wavefunction in
scalar spherical harmonics on $S^8$:
\begin{equation}
\widetilde\psi_a(\mathbf n)
=
\sum_{\ell,m}c_{\ell m,a}Y_{\ell m}(\mathbf n),
\qquad
-\nabla_{S^8}^2Y_{\ell m}
=
\ell(\ell+7)Y_{\ell m}.
\end{equation}
The label $\ell$ specifies a symmetric traceless
SO(9) tensor representation of rank $\ell$, and $m$
labels its components.\footnote{Scalar spherical harmonics are restrictions to the unit
sphere of homogeneous harmonic polynomials
$P_\ell = T_{i_1\cdots i_\ell}x^{i_1}\cdots x^{i_\ell}$.
Since $x^i$ commute, only the symmetric part of $T$
contributes; the harmonic condition $\nabla^2P_\ell=0$ requires
$T$ to be traceless.}

Only even $\ell$ occur. To see this, the residual SU(2)
gauge transformation exchanging the two D0-branes sends
$\mathbf n\to-\mathbf n$ and
$\theta_\alpha\to-\theta_\alpha$.
The latter acts on the polarizations as fermion parity
$\Pi$, equal to $+1$ on
$\mathbf{44}\oplus\mathbf{84}$ and $-1$ on
$\mathbf{128}$. Gauge invariance requires
$\psi(-\mathbf n)=\Pi\psi(\mathbf n)$.
Using $W(-\mathbf n)=\Pi W(\mathbf n)$ gives
\begin{equation}
\widetilde\psi(-\mathbf n)=\widetilde\psi(\mathbf n).
\end{equation}
Since $Y_{\ell m}(-\mathbf n)=(-1)^\ell
Y_{\ell m}(\mathbf n)$, the transformed wavefunction
contains only even $\ell=0,2,4,\ldots$. This means that the allowed irreps of $R$ for scattering states are
\begin{align}\label{eq:orbital}
\text{orbital SO(9) angular momentum} \in \{\mathbf{1},\mathbf{44},\mathbf{450},\ldots\}.
\end{align}
Because $f$ depends only on $\mathbf n'\cdot\mathbf n$,
the spherical harmonics diagonalize its action:
\begin{equation}\label{eq:smatrix-angular-eigenvalue}
\int_{S^8}d\Omega_{\mathbf n}\,
f(E,\mathbf n'\cdot\mathbf n)Y_{\ell m}(\mathbf n)
=
S_\ell(E)Y_{\ell m}(\mathbf n').
\end{equation}
Consequently,
\begin{equation}
c_{\ell m,a}^{\mathrm{out}}
=
S_\ell(E)c_{\ell m,a}^{\mathrm{in}}.
\end{equation}
Elasticity requires equal incoming and outgoing fluxes
in each scattering eigenchannel. Therefore
$|S_\ell(E)|=1$, and we define the real phase shift by
\begin{equation}\label{eq:smatrix-phase-definition}
S_\ell(E)=e^{2i\delta_\ell(E)},
\qquad
\delta_\ell(E)\in\mathbb R\pmod{\pi}.
\end{equation}
The phase shift is the phase of the outgoing wave
relative to the incoming wave, with the free-wave
convention removed. 

One can compute $\delta_\ell(E)$ in a 1-loop approximation, using the effective Lagrangian $\mathcal{L}_{\rm int}=\frac{15}{16}v^4/r^7$ (and its supersymmetric completion) \cite{Douglas:1996yp,Plefka:1998ed}; this gives
\begin{align}
    \tan\delta_\ell(E)
    &= a_\ell E^{9/2} + \cdots,
    \qquad
    a_\ell = \frac{4\,\ell!}{(\ell+6)!},
    \qquad \ell=0,2,4,\ldots .
    \label{eq:1loop-phase}
\end{align}
At two loops, the effective Lagrangian contains a term proportional
to $v^6/r^{14}$ \cite{Becker:1997xw}. This interaction and the
iteration of the one-loop interaction both contribute to the
coefficient $b_\ell$ of $E^9$. Thus, for even $\ell\geq 2$ we have\footnote{For $\ell=0$, a short-distance term of order $E^{15/2}$ is allowed
before the $E^9$ term.}
$\tan\delta_\ell(E)
= a_\ell E^{9/2} + b_\ell E^9 + \cdots .$
At large $\ell$ we can also estimate how the range of validity of
\eqref{eq:1loop-phase} grows with angular momentum. Let $b$ denote the semiclassical impact parameter. 
Then the derivative expansion is controlled when $q\ll 1$, where $q \propto v/b^2 \propto E^{3/2}/(\ell+\tfrac72)^2$.
The loop expansion is controlled when $g_{\rm YM}^2/b^3 \propto q/(\ell+\tfrac72) \ll 1$.
Thus both expansions are under control when
\begin{align} \label{eq:convergenceRegion}
    E \ll \bigl(\ell+\tfrac72\bigr)^{4/3}.
\end{align}
In figure \ref{fig:scatteringPhase}, we compare this Born-Oppenheimer prediction against the extracted S-matrix and see agreement in this regime.

The free radial equation has solutions expressed in spherical Bessel functions of
order $L=\ell+3$. At large separation, the radial wavefunction is a linear combination %
\begin{equation}\label{eq:smatrix-radial-in-out}
u_\ell(r)\sim
\frac{C_\ell pr}{2}
\left[
e^{2i\delta_\ell(E)}h_{\ell+3}^{(1)}(pr)
+
h_{\ell+3}^{(2)}(pr)
\right],
\end{equation}
where $h_L^{(1)}=j_L+i y_L$ and $h_L^{(2)}=j_L-i y_L$
are the outgoing and incoming spherical Hankel functions.
At large $pr$ this reduces to
\begin{equation}
u_\ell(r)\sim
C_\ell e^{i\delta_\ell(E)}
\sin\!\left(
pr-\frac{(\ell+3)\pi}{2}+\delta_\ell(E)
\right).
\end{equation}

Finally, one can also invert the relation \eqref{eq:smatrix-angular-eigenvalue} to express $f$ in terms of $\delta_\ell$: %
\begin{equation}\label{eq:smatrix-partial-expansion}
f(E,\cos\theta)
=
\frac{1}{7\Omega_8}
\sum_{\ell=0,2,4,\ldots}
(2\ell+7)C_\ell^{7/2}(\cos\theta)e^{2i\delta_\ell(E)},
\qquad
\Omega_8=\frac{32\pi^4}{105},
\end{equation}
The functions $C_\ell^{7/2}$ are Gegenbauer polynomials. Here $f$ is the kernel of the full S-operator, including
the identity contribution.
The connected kernel is obtained by replacing
$e^{2i\delta_\ell(E)}$ with
$e^{2i\delta_\ell(E)}-1$.

\subsection{Relation to the total SO(9) sectors}
\label{sec:smatrix-total-sectors}
The partial-wave label $\ell$ is not the total
SO(9) representation $J$ used in the numerical
calculation. In the transformed basis, the scattering space at fixed energy is
\begin{equation}\label{eq:scattering-multiplets}
\mathcal H_E^{\mathrm{scatt}}
\simeq\bigoplus_{\ell=0,2,4,\ldots}S\otimes[\ell],
\qquad [\ell]=(\ell,0,0,0).
\end{equation}
Here $S$ is defined in \eqref{eq:supermultiplet}, and the rank-$\ell$
symmetric traceless representation $[\ell]$ is the superspin $R$.
All states in each $S\otimes[\ell]$ multiplet have the same
scattering phase $\delta_\ell(E)$.

For a total SO(9) singlet, only the contraction
$\mathbf{44}\otimes[2]\supset\mathbf1$ contributes.
Explicitly, the asymptotic state is
$
|k;J=\mathbf1\rangle
\propto
\int_{S^8}d\Omega_{\mathbf n}\,
\left(n_i n_j-\frac{\delta_{ij}}9\right)
|k\mathbf n;ij\rangle.$
Its angular-polarization tensor is unchanged by
$W(\mathbf n)$, so the scalar partial-wave label remains
$\ell=2$. The singlet phase is consequently
$
\delta_{\mathrm{sing}}(E)=\delta_2(E)\pmod{\pi}.
$

For the other even-rank symmetric traceless total
representations $J=[m]$ considered here, only the
polarization $\mathbf{44}$ contributes.
For $m\geq2$, the allowed scalar partial waves are
$\ell=m-2,m,m+2$.
The sectors of interest therefore contain
\begin{equation}\label{eq:smatrix-sector-waves}
\begin{array}{c|ccccc}
J & [0] & [2] & [4] & [6] & [8] \\\hline
\ell & 2 & 0,2,4 & 2,4,6 & 4,6,8 & 6,8,10
\end{array}
\end{equation}
Several orbital channels can have the same total SO(9) irrep $J$.
Scattering preserves $J$ but does not necessarily preserve orbital angular momentum, so the states
that acquire a single scattering phase can be linear combinations
of the orbital basis states $\Phi_{J,c}$.

At fixed $J$, we use the basis
$\Phi^{JM}_{\ell}(\mathbf n)=\sum_{m,a}
\langle \ell m;\mathbf{44}\,a\mid JM\rangle\,
Y_{\ell m}(\mathbf n)\,|a\rangle$,
where $\ell$ is the orbital angular momentum. The matrix elements of $W(\mathbf n)$ in this basis define the change of basis matrix
\begin{equation}
(V_J)_{\ell'\ell}
=\int_{S^8}d\Omega_{\mathbf n}\,
\Phi^{JM}_{\ell'}(\mathbf n)^\dagger
W(\mathbf n)\Phi^{JM}_{\ell}(\mathbf n)
\end{equation}
to states that scatter with a definite phase $\delta_\ell(E)$.
The fixed $J$ S-matrix is then
\begin{equation}\label{eq:smatrix-sector-S}
S_J(E)
=
V_J\,
\operatorname{diag}\!\left(
e^{2i\delta_{\ell_1}(E)},
e^{2i\delta_{\ell_2}(E)},\ldots
\right)V_J^T.
\end{equation}
For $J=\mathbf{44}$, its three eigenphases are
$\delta_0,\delta_2,\delta_4$.
The same $\delta_2$ also appears in the $J=[0]$ (singlet) and
$J=[4]$ sectors, as required by supersymmetry.
Extracting $\delta_2$ from different sectors therefore provides a non-trivial consistency check on the numerical calculation.

\subsection{Extracting the phase shift from BFSS in a box}\label{sec:smatrix-box}

\begin{figure}[htbp]
\centering
\includegraphics[width=0.7\textwidth]{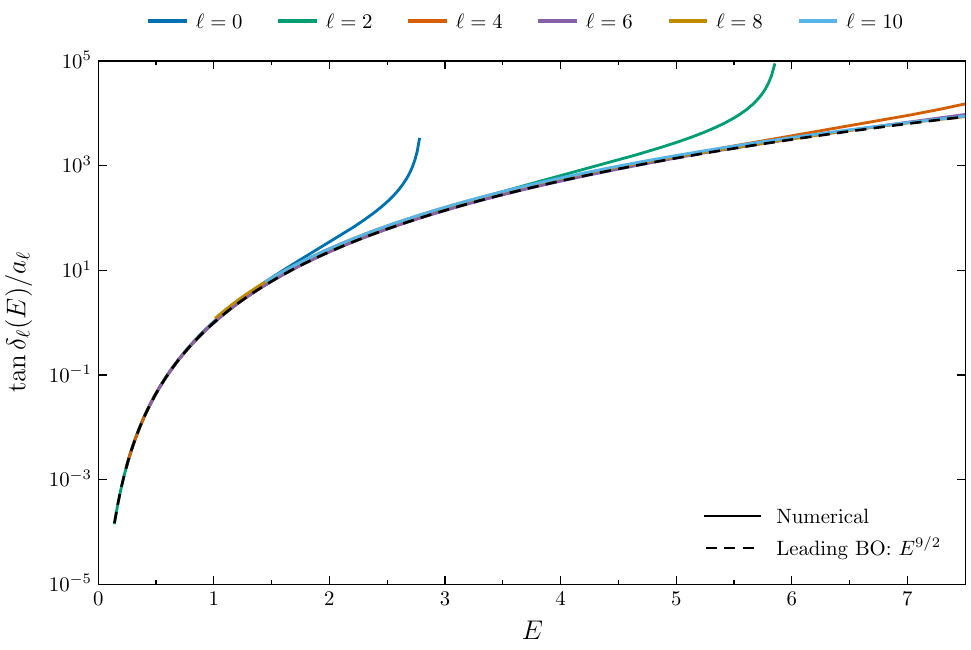}
\caption{Comparison of the scattering phase $\delta_\ell(E)$ with the Born-Oppenheimer approximation. We compare against the 1-loop term extracted from $\mathcal{L} \supset 15 v^4/ (16 r^7)$, see equation \eqref{eq:1loop-phase}. We see good convergence as $\ell$ increases and at low energies, as predicted by \eqref{eq:convergenceRegion}.}
\label{fig:scatteringPhase}
\end{figure}

\begin{figure}[htbp]
\centering
\includegraphics[width=\textwidth]{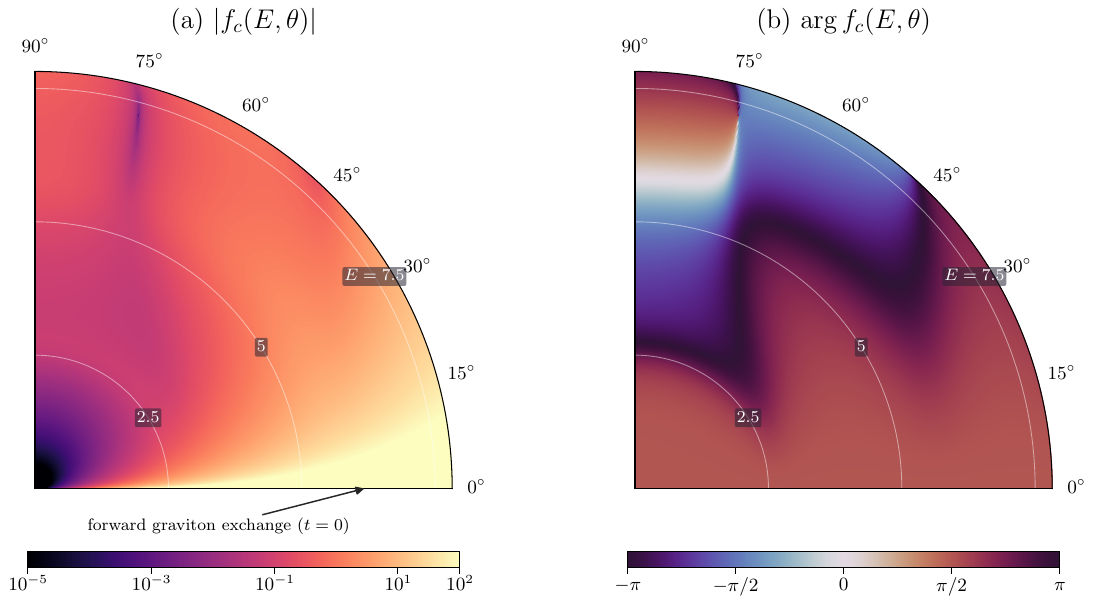}
\caption{Magnitude (a) and phase (b) of the connected scattering
amplitude $f_c(E,\theta)$.
Radius denotes the energy $E=p^2$ and angle denotes the scattering
angle $\theta$. At any
fixed $E>0$, the momentum transfer $t=-2E(1-\cos\theta)$ vanishes
as $\theta\to0$ (forward limit), giving the $t$-channel graviton pole.
The reconstruction uses the phase shifts extracted from the L\"uscher method for $\ell=0,2,4,\cdots, 10$
and the leading Born-Oppenheimer prediction for higher even partial waves.
Exchange symmetry relates $\theta$ to $\pi-\theta$, so only 
$0\leq\theta\leq\pi/2$ is shown.}
\label{fig:amplitude}
\end{figure}

We now determine the phases using L\"uscher's method~\cite{Luscher:1990ux}. 
A closely related approach was applied to supersymmetric quantum
mechanics to extract S-matrices and metastable resonances
in~\cite{Balthazar:2016utu}.
Let us first recall this method for a 1D non-relativistic scattering problem for a short range potential. The idea is to modify the problem by placing the particle in a box. In the asymptotic region, the general solution of Schr\"odinger's equation is $u(r) \propto   \sin (k r + \delta(k))$. Then, placing a hard wall at $r={R_c}$ gives 
\begin{align}
    u({R_c}) = 0 \Rightarrow k {R_c} +\delta(k) = n \pi .
\end{align}
Therefore the allowed energies are $\delta(E_n({R_c})) =n \pi - k_n {R_c}$. If we measure the spectrum at some value of ${R_c}$, we determine many $E_n$ and therefore the value of the phase shift $\delta(E_n({R_c}))$ at many discrete values. By varying ${R_c}$ continuously, we can measure $\delta(E)$.

Now let us turn to the BFSS problem.
We first derive the relation between the energies for the theory in a spherically symmetric box and the scattering phase shift, treating the relative motion as free near the wall and at larger separations\footnote{After including the wall and interaction-tail corrections, we check that a common set of phase shifts describes the spectra at different box sizes. We also test the fitted phases against scattering levels in additional boxes and in an independent SO(9) sector excluded from the fit.}.
{In each total SO(9) representation $J$, we solve the Schr\"odinger equation with a wall at $r=R_c$, or equivalently $s_1=R_c/\sqrt2$:}
\begin{equation}\label{eq:smatrix-box-schrodinger}
H\Psi_n^{(J)}=E_n^{(J)}({R_c})\Psi_n^{(J)},\qquad
\Psi_n^{(J)}\big|_{{s_1=R_c/\sqrt2}}=0.
\end{equation}
We find successive levels by minimizing the expectation value of the Hamiltonian\footnote{Note that this is a second order formulation, which differs from the construction in  Section \ref{sec:ground} where we use the first order formulation since we want to study the BPS state.} in the box,
with the additional requirement that successive states be orthogonal to the lower energy states. This gives a discrete set of energy levels. Most of these energy levels are associated with scattering states; a small subset will be associated to bound states which we will discuss in Section \ref{sec:bdstates}.

Let us now consider what scattering states are associated with energies in each overall SO(9) irrep $J$. For simplicity let us start with the trivial irrep.
For a total SO(9) singlet, the $\mathbf{44}$ polarization
contracts with an orbital $\ell=2$. The radial solution
therefore uses spherical Bessels of order 5, and its scattering phase is
$\delta_2(E)$. Imposing ${u_2(R_c)=0}$ gives
\begin{equation}\label{eq:smatrix-singlet-box}
{\tan\delta_2(E_n)=\frac{j_5(p_nR_c)}{y_5(p_nR_c)},\qquad p_n=\sqrt{E_n}.}
\end{equation}
Here $p_n$ is conjugate to $r$. %
Each singlet scattering level $E_n$ gives one value of $\delta_2(E_n)$; we follow the phase continuously as the energy changes.

Now consider $J=m$, with $m=2,4,6,8$. There are three ways to combine orbital angular momentum with the $\mathbf{44}$ polarization to obtain this $J$, with $\ell \in  \{m-2, m, m+2 \}$. 
An incoming wave in any one of these three
channels can scatter into an outgoing superposition of all three.
The hard-wall condition must therefore be imposed on all three radial
components.
Using the change of basis $V_J$ of \eqref{eq:smatrix-sector-S}, the
analogue of \eqref{eq:smatrix-singlet-box} becomes
\begin{equation}\label{eq:smatrix-coupled-box}
\det\!\left[D_J({pR_c})-K_J(E)\right]=0,
\end{equation}
where $D_J$ contains the spherical Bessels for the three
orbital angular momenta,
\begin{equation*}
D_J(z)=\operatorname{diag}\!\left(
\frac{j_{m+1}(z)}{y_{m+1}(z)},
\frac{j_{m+3}(z)}{y_{m+3}(z)},
\frac{j_{m+5}(z)}{y_{m+5}(z)}
\right)
\end{equation*}
and the matrix $K_J$ is
\begin{equation}\label{eq:smatrix-K}
K_J(E)=V_J\,\operatorname{diag}\!\left(
\tan\delta_{m-2}(E),\,\tan\delta_m(E),\,\tan\delta_{m+2}(E)
\right)V_J^T .
\end{equation}
We can reconstruct the scattering matrix in \eqref{eq:smatrix-sector-S} via $S_J = (1 + \i K_J) (1-\i K_J)^{-1}$. In practice, to extract these phases as a function of the energies, we vary the radius of the box $R_c$. We show the results for the scattering phase in Figure \ref{fig:scatteringPhase}, where we also compare against the Born-Oppenheimer approximation.

\section{Excited bound states \label{sec:bdstates}}

Recall from Section~\ref{sec:supermultiplet} states for $E>0$ are organized into supermultiplets. These supermultiplets are labeled by two quantum numbers $S \otimes R$. $R$ is the SO(9) irrep of the top state; acting with the supercharges on the top state fills out an additional 256 dimensional factor $S$ according to \eqref{eq:supermultiplet}. This supermultiplet structure applies to both bound states and scattering states.

We will now apply this in the following way. Suppose one finds a state that is an overall SO(9) singlet. SUSY predicts that it is a member of three possible families of supermultiplets: $\mathbf{44} \otimes \mathbf{44}$, $\mathbf{84} \otimes \mathbf{84}$ or $\mathbf{128} \otimes \mathbf{128}$. We will argue that in the latter two cases, such a state must be a {\it bound} state, not a scattering state.
Then we prove that infinitely many states exist with these quantum numbers, which implies that there are infinitely many bound states that are SO(9) singlets.

The proof shows that such states exist, but does not compute their eigen-energies. We report on such energies and eigenfunctions in Section~\ref{sec:numerical}. We also find evidence of some resonances (or metastable states).

\subsection{Existence}

 Recall that the scattering supermultiplets in \eqref{eq:scattering-multiplets} have $R=[\ell]$ with even $\ell$. 
 This is because the partial waves responsible for the orbital angular momentum \eqref{eq:orbital} are associated with symmetric traceless tensors of even rank.
 To make a total SO(9) singlet, the only possibility is to contract $R=[2]=\mathbf{44}$ with $S=\mathbf{44}$. This means that scattering states that are an overall SO(9) singlet cannot carry $\mathbf{84} \otimes\mathbf{84} $ or $\mathbf{128} \otimes\mathbf{128}$ quantum numbers. Hence we immediately conclude that an eigenstate of finite energy $E>0$ must be a bound state if it carries quantum numbers $\mathbf{84} \otimes \mathbf{84}$ or $\mathbf{128} \otimes \mathbf{128}$.

Note that the above argument is specific to SU(2) BFSS. For SU($N$) with
$N\ge 3$, scattering states can contain $\ge 3$ 
clusters. $R$ receives contributions from both relative
orbital motion and additional spins (polarization states of the additional fermions), and is
thus no longer restricted to symmetric traceless tensors.
Indeed, already in SU(3), the three-D0 scattering channel
contains total SO(9) singlets with $R=\mathbf{84}$ or
$\mathbf{128}$. The quantum numbers therefore do not automatically exclude scattering states at higher $N$.

To prove the existence of a single positive-energy bound state, it remains to show that the protected singlet sector is nonempty\footnote{We do not need to construct an energy eigenfunction: any  normalizable state in this sector suffices, since its positive-energy spectrum consists of bound-state eigenvalues.}. For this, we consider the operator $\mathcal{O} = \Tr (X^{I} X^{J} X^K \psi_\alpha  \psi_\beta)  \gamma^{IJK}_{\alpha \beta}$ and perturb the ground state by this operator:
\begin{align}
    \ket{\mathcal{O}} = \Tr (X^{I} X^{J} X^K \psi_\alpha  \psi_\beta)  \gamma^{IJK}_{\alpha \beta} \ket\Omega.
 \end{align}
 As before $\ket\Omega$ is the normalizable ground state. 
To argue that $\ket\O$ is normalizable, note that $\mathcal O$ is cubic in the bosonic coordinates. Since the ground state has the power law tail $r^{-9}$ the norm of $\ket\O$ is
\begin{equation}
    \bra{\O}\ket{\O} = \int^\infty \! r^8\,r^6\,r^{-18}\,dr<\infty.
\end{equation}
Since $\gamma^{IJK}$ is an antisymmetric tensor in its three bosonic indices, the form of the state strongly suggests that it has an overlap in the $\mathbf{84} \otimes \mathbf{84}$ irrep. We argue for this more carefully in Appendix \ref{app:boundState}. Furthermore, we show that variants of this state have an energy that grows without bound.

\subsection{Positive energy numerical solutions \label{sec:numerical}}

\begin{figure}[!htbp]
    \centering
    \includegraphics[width=\textwidth]{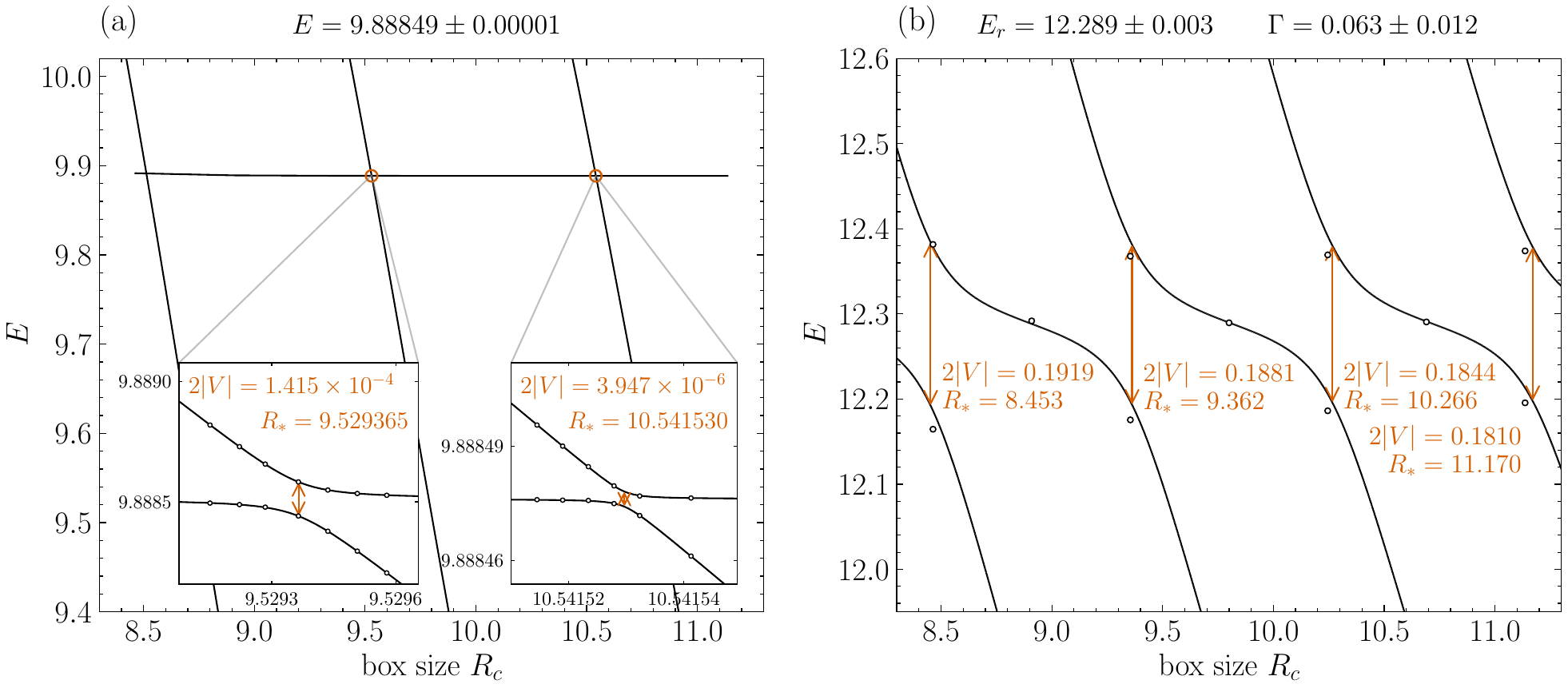}
    \caption{Finding  bound states and resonances with the L\"uscher method. The inset panels depict the eigenvalue repulsion as scattering levels cross the localized ones. The left panel shows the splitting decreases by a factor of $\sim 30$ between the two crossings, a signature of a bound state, while for the narrow resonance in the right panel the splitting remains stable. The bound state appears in the {\bf 128} column in {Figure~\ref{fig:spectrum-family}} whereas the resonance appears in the {\bf 44} column. }
    \label{fig:7849_bound_state}
\end{figure}

We use the same finite-box spectrum used in the extraction of the S-matrix for L\"uscher's method. This spectrum shows clear evidence of both bound states and metastable states or
resonances. %
As we increase the IR cutoff $R_c$, the scattering levels decrease roughly as $\propto 1/R_c^2$ until they encounter a bound state level, whose energy is approximately independent of $R_c$, see Figure~\ref{fig:7849_bound_state}. Near this encounter, eigenvalue repulsion kicks in and a two-level approximation gives
\begin{equation}\label{eq:avoided-crossing-width}
    {\Delta E(R_c)\simeq
    \sqrt{a^2(R_c-R_*)^2+4|V_*|^2},
    \qquad
    \Gamma\simeq\frac{2\pi|V_*|^2}{D}}
\end{equation}
{Here $R_*$ is the radius at which crossing nearly occurs, $V_*$ is the mixing matrix element at the crossing, and $D$ is the local spacing of the background scattering levels.}
For a resonance, the inferred width approaches a
nonzero value as the infrared cutoff $R_c \to\infty$, whereas for a stable bound state $\Gamma \to 0$ as $R_c \to \infty$.

Figure~\ref{fig:7849_bound_state} illustrates the distinction.
For the bound state in the left panel, the minimum splitting
decreases by a large factor between successive
encounters with scattering levels. This supports its
interpretation as a stable bound state.
An even stronger piece of evidence that it is a bound state is its supermultiplet assignment (the procedure of assigning a state to a supermultiplet is described below). We found that this state is consistent with an $R={\mathbf{128}}$ supermultiplet; as reviewed above there are no scattering states with $R={\mathbf{128}}$ or $R={\mathbf{84}}$.
The right panel shows a metastable state: its scattering
phase is described by a Breit--Wigner resonance with
$E_r=12.289\pm0.003$ and $\Gamma=0.063\pm0.012$.
{The resolved resonance energies and widths are summarized in Table~\ref{tab:resonance-numerics}.}

\begin{table}[!htbp]
\centering
{
\setlength{\tabcolsep}{7pt}
\begin{tabular}{ccc}
\toprule
$R=[\ell]$ & $E_r\pm\delta E_r$ & $\Gamma\pm\delta\Gamma$ \\
\midrule
$\mathbf{1}$   & $2.6068\pm0.00025$       & $0.7388\pm0.00063$ \\
$\mathbf{44}$  & $5.6796\pm0.0026$        & $1.4204\pm0.011$   \\
$\mathbf{44}$  & $8.2126\pm0.01$          & $0.8834\pm0.01$    \\
$\mathbf{1}$   & $8.5355\pm0.03$          & $0.7471\pm0.064$   \\
$\mathbf{450}$ & $8.5753\pm0.0067$        & $0.6887\pm0.024$   \\
$\mathbf{44}$  & $(10.6381\pm0.0069)$     & $0.5543\pm0.017$   \\
$\mathbf{44}$  & $(11.7507\pm0.059)^{a}$  & $0.4289\pm0.14$    \\
$\mathbf{44}$  & $(12.2885\pm0.0031)^{b}$ & $0.0627\pm0.012$   \\
\bottomrule
\end{tabular}
\caption{Resonance energies and widths.
The central values are S-matrix poles (obtained by fitting), and
the quoted uncertainties are the largest changes across
alternative fits.
${}^{a}$Central values from a Breit--Wigner fit to the
singlet-sector phase; uncertainties are estimated from variation
across alternative S-matrix pole fits.
${}^{b}$Breit--Wigner fit to 17 readings from seven singlet
boxes, with statistical fit errors.
The quoted uncertainties do not include all systematic effects.}
\label{tab:resonance-numerics}
}
\end{table}

\begin{figure}[!htbp]
    \centering
    \includegraphics[width=\textwidth]{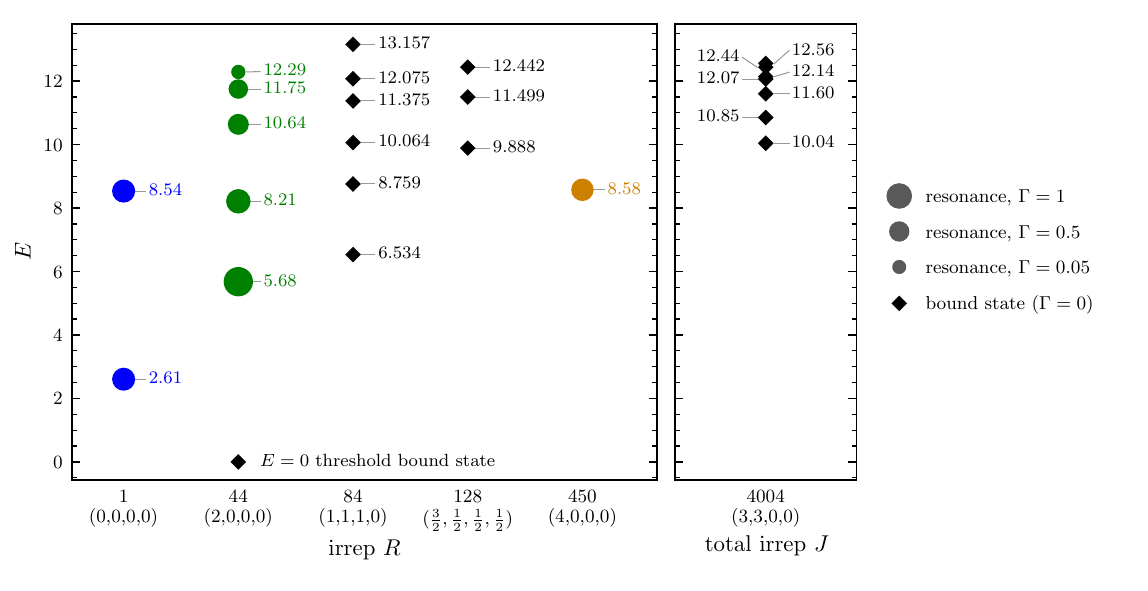}
    \caption{Bound states and resonances resolved numerically. {\it Left}: levels grouped by the superspin
    $R$ of their multiplet. Circles denote resonances, with larger
    circles indicating larger decay widths $\Gamma$; diamonds denote
    bound states. The bound states in the $\mathbf{84}$ and
    $\mathbf{128}$ columns were found in the total SO(9) singlet
    sector. The threshold bound state at $E=0$ does not carry a superspin label $R$ but its asymptotic tail is in the orbital $\ell=2$ or $R=\mathbf{44}$ irrep. {\it Right}: bound states found at fixed total SO(9)
    irrep $J=(3,3,0,0)$, of dimension $4004$. There are no scattering states with this $J$.} %
    \label{fig:spectrum-family}
\end{figure}

We find several positive-energy bound states, summarized in
Figure~\ref{fig:spectrum-family}. In the SO(9) singlet sector,
we resolve 9 positive-energy levels associated to 6 different multiplets with
$R=\mathbf{84}$ and 3 with $R=\mathbf{128}$.
Here we count each supermultiplet once, using its singlet
representative. 
{Table~\ref{tab:singlet-bound-numerics} gives the energies, measured convergence changes, and supercharge diagnostics for these nine states.}

\begin{table}[!htbp]
\centering
{
\begin{tabular}{cccc}
\toprule
$n$ & $R$ & $E\pm\delta E$ & $w$ \\
\midrule
1 & $\mathbf{84}$  & $6.533789\pm0.00009$  & $0.9995$ \\
2 & $\mathbf{84}$  & $8.758748\pm0.00002$  & $0.9992$ \\
3 & $\mathbf{128}$ & $9.888492\pm0.00002$  & $0.9979$ \\
4 & $\mathbf{84}$  & $10.063992\pm0.00003$ & $0.9987$ \\
5 & $\mathbf{84}$  & $11.374740\pm0.000004$ & $0.9992$ \\
6 & $\mathbf{128}$ & $11.498879\pm0.00003$ & $0.9984$ \\
7 & $\mathbf{84}$  & $12.075276\pm0.00002$ & $0.9996$ \\
8 & $\mathbf{128}$ & $12.441947\pm0.00002$ & $0.9093$ \\
9 & $\mathbf{84}$  & $13.156824\pm0.00010$ & $0.9877$ \\
\bottomrule
\end{tabular}
\caption{Positive-energy bound states. Energy uncertainties are a rough attempt to quantify systematic errors due to box-size dependence and grid resolution.
The supermultiplet labels $R$ are inferred from the measured
weight $w$ in the matching $\mathbf{84}$ or $\mathbf{128}$
component of $S$. These weights retain finite-resolution effects; their deviation from $w=1$ is a way to quantify how well each energy level has converged.}
\label{tab:singlet-bound-numerics}
}
\end{table}
We now describe the other checks on the lowest such
state in detail. The lowest positive-energy bound state that we find is an SO(9) singlet with energy $E\simeq 6.534$, which we will assign to $R=\mathbf{84}$.
We checked that its probability density is localized well in the interior of the box,
and its energy becomes insensitive to the IR cutoff\footnote{In addition to a hard wall cutoff, we tried a harmonic trap $H' = H_\mathrm{BFSS} + \hf \mu^2 \Tr (X^I)^2$}. We also plotted its radial probability distribution for various values of ${R_c}$ and observed only minor variation at large $R_c$. %
\begin{figure}[!htbp]
    \centering
    \includegraphics[width=\linewidth]{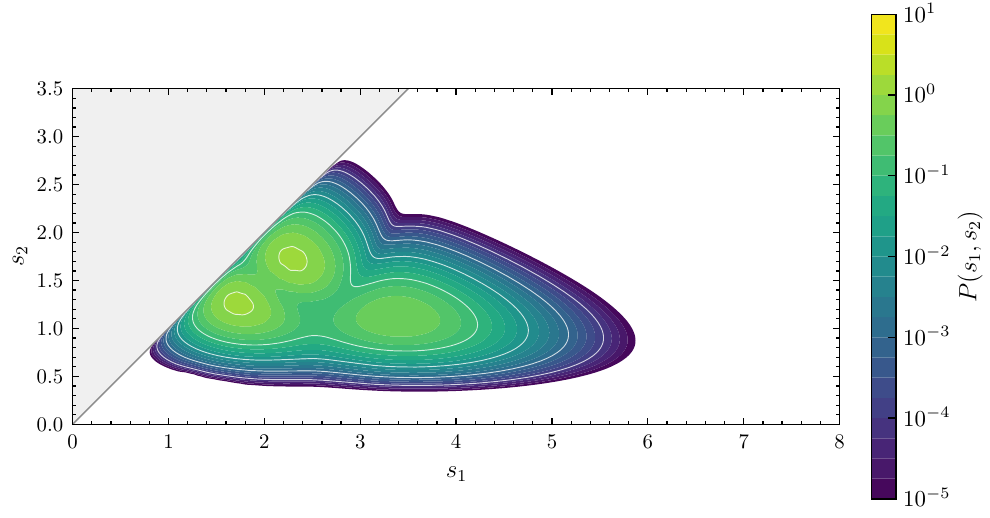}
    \caption{The probability density $P(s_1,s_2)$ of the lowest
positive-energy SO(9)-singlet bound state, with $E\simeq 6.534$, as in Figure~\ref{fig:gs-s1-s2}. The distribution exhibits three broad peaks. Unlike the ground state $E=0$ depicted in Figure~\ref{fig:gs-s1-s2}, the $E>0$ bound states have no power law tail; the density decreases rapidly for large $s_1$. %
}
    \label{fig:boundState_s1s2}
\end{figure}

We tested that the bound state comes in the expected supersymmetry multiplet\footnote{Since our ``finite box'' boundary conditions break SUSY, this statement is approximately true when $R_c$ is large.}.  Starting from the numerical state that is an SO(9) singlet,
we apply chains of normalized supercharges and use the
resulting overlaps to extract its weights in the SO(9) components of $S$
in \eqref{eq:supermultiplet}. We find a weight of approximately
$0.9995$ in the $\mathbf{84}$ component of $S$.
Given that the state is an overall SO(9) singlet, this implies
$R=\mathbf{84}$ and we identify the supermultiplet as
\begin{equation} \label{decomposition}
    S\otimes\mathbf{84} = \mathbf{1} + \mathbf{36} + 0(\mathbf{9}) + \mathbf{44} + \cdots .
\end{equation}
Here we have omitted other irreps that we did not numerically study.
We compared this to the states that we found in other low-lying irreps of SO(9) and looked for states with the same energy.
At the same box size, we found a state with
\(E=6.5339241\) in the singlet sector and \(E=6.5339309\) in the
\(\mathbf{44}\) sector. As we refined the grid, the splitting decreases.
We found another bound state in the ${\bf 36}$ sector, and no bound state in the vector ${\bf 9}$ state. These facts are all consistent with \eqref{decomposition}, supporting the identification of all of these states as members of a common supermultiplet. As explained above, $R= \mathbf{84}$ is absent from the two-D0
scattering continuum, so the corresponding exact multiplet
is stable.

For higher SO(9) singlet bound states, we also check the supermultiplet
assignment using the supercharges, follow the energy as the box
size changes, and examine mixing with scattering levels.
We distinguish the $R=\mathbf{84}$ and $R=\mathbf{128}$ assignments
by looking for levels at the same energy in other SO(9)
representations. The former predicts a single level in the
$J=\mathbf{36}$ sector and 0 in $J=\mathbf{9}$, whereas the
latter predicts 2 degenerate levels in $J=\mathbf{36}$ and a single level in
$J=\mathbf{9}$.

We also find bound states in the total SO(9) representation
$J=(3,3,0,0)$, see the right panel of Figure~\ref{fig:spectrum-family}. This irrep of $J$ also contains no SU(2) scattering states.  
Indeed, the scattering multiplets in \eqref{eq:scattering-multiplets}
have the form $S\otimes(\ell,0,0,0)$ with even $\ell$. 
Every irrep that appears in the decomposition has second highest-weight component at most $2$, excluding $J=(3,3,0,0)$.
This is essentially the mechanism explained in the OpenAI proof \cite{OpenAIMath:2026bfss270}.

\section{Discussion}
In this work we solved for the ground state wavefunction for SU(2) BFSS. 
We also solved for the S-matrix and found bound state wavefunctions. 
We did not exhaustively search for the bound state wavefunctions; it would be interesting to do this in future work and try to understand analytically aspects of the bound state spectrum $E_n$ in various irreps of SO(9).

Working in gauge-invariant and SO(9)-invariant coordinates was crucial to make the numerics for SU(2) practical. For larger values of $N$ numerics will become rapidly more difficult. 
To make progress, one could imagine searching for more efficient variational ansatz. As we will report in \cite{BajajEtAl:BFSSBootstrap}, bootstrap islands can be obtained for SU(3) BFSS so any proposed ansatz would benefit from rigorous checks.

We also found resonances, or metastable states. Of course, there is no dimensionless parameter in SU(2) BFSS that can make these metastable states parametrically long lived. However, holography suggests that black holes in the 't Hooft/type IIA regime are exponentially long lived with a lifetime $\Gamma^{-1} \sim e^{\Delta S(E)}$, where the change in entropy of emitting a D0-brane is $\Delta S \propto N$; see Appendix A of \cite{Lin:2013jra}. Therefore, it is tempting to speculate that these might be finite $N$ remnants of such ``black hole''-like resonances.
It would be interesting to study the wavefunction of these metastable states using the L\"uscher method; perhaps one could distinguish resonances with black-hole features from other kinds of resonances, inspired by the fortuity/monotonous discussion \cite{Chang:2024zqi,Chen:2024oqv,Chang:2024lxt}.

Note that the numerical solution of the bound state wavefunction(s) is a first step towards the non-perturbative construction of the S-matrix of larger $N$ BFSS theories. In higher $N$ BFSS theories, we may always consider a scattering state which is formed by considering $2\! \times \!2$ block(s) that are in some bound state of SU(2) BFSS. This bound state may be a zero energy or $E>0$ bound state. In addition, we expect that the power law tails of higher $N$ BFSS to similarly contain the ground states of lower $N$ BFSS, see \cite{Lin:2014wka}.

The fact that many bound states with various different representations of SO(9) exist in the SU(2) theory therefore implies that the scattering states of SU($N$) are considerably richer than previously believed. Presumably in the low energy limit relevant for the BFSS conjecture, these scattering states will not be relevant since they involve energies much larger than the energies involved in the BFSS limit. Nevertheless, scattering at finite $N$ or in the planar 't Hooft regime is of interest (say probe D0-brane scattering in the extremal D0 background), apart from the BFSS conjecture. This deserves further study.

\section*{Acknowledgements}
We thank Alexander Michel, Jaroslav Scheinpflug, and Xi Yin for discussions and for sharing their manuscript with us.
We thank Vasily Dommes, David Simmons-Duffin, and Zechuan Zheng for collaboration on related work and useful discussions. We also benefited from discussions with Igor Klebanov, Juan Maldacena and Douglas Stanford.

\section*{AI statement}
Generative AI models (6.1 Sol, Astra, Fable, Opus 5.5) were used to write code, check analytical derivations, and for manuscript revisions. KB acknowledges OpenAI's academic research program for providing access to Codex.

\appendix
\section{The fermionic Hilbert space}\label{app:fermions}

The fermions rotate under SO(9) and SO(3) with the generators
\begin{equation}\label{eq:gens}
\Sigma_{IJ}=\frac14\,\psi_{\alpha c}\gamma^{IJ}_{\alpha\beta}\psi_{\beta c},\qquad\Lambda_{ab}=\psi_{\alpha a}\psi_{\alpha b}.
\end{equation}
At a point of the slice the rotations of the directions $I=4,\dots,9$, a Spin(6)\,$\simeq$\,SU(4), do not move $X$.
Under Spin(3)\,$\times$\,Spin(6) the spinor index splits as $\mathbf{16}=(\mathbf2,\mathbf4)\oplus(\mathbf2,\bar{\mathbf4})$,
so the 48 Majorana fermions pair into 24 complex ones $c^i_A$, with $i$ an SU(4) index and $A=1,\dots,6$ running over the
Spin(3) doublet and the three colours. The SU(4)-invariant states are built from
\begin{equation}
\epsilon_{ijkl}\,c^{i\dagger}_Ac^{j\dagger}_Bc^{k\dagger}_Cc^{l\dagger}_D\ket{0},
\end{equation}
so they contain $0,4,\dots,24$ fermions. A rotation by $\pi$ in the plane of the directions
$a,b\leq3$ flips the signs of $X^a$ and $X^b$, and the gauge rotation by $\pi$ in the same colour plane flips them back.
Their product also leaves the point fixed, so
\begin{equation}
e^{\pi(\Sigma_{ab}+\Lambda_{ab})}\,\Omega(s)=\Omega(s),
\end{equation}
and $\mathfrak F$ is the subspace of SU(4)-invariant states that obey these conditions.
To count its dimension, first let $\mathcal V=\mathcal H_f^{\mathrm{SU}(4)}$.
The numbers of SU(4) singlets in the sectors with $0,4,8,12,16,20,24$ fermions are
$1,126,1764,4116,1764,126,1$, respectively, giving $\dim\mathcal V=7898$.
On $\mathcal V$, the combined rotations $G_{ab}=e^{\pi(\Sigma_{ab}+\Lambda_{ab})}$
generate $\mathbb Z_2\times\mathbb Z_2$, so only two of the three conditions are independent.
The projector onto their common invariant subspace is
\begin{equation}
P_{\mathfrak F}=\frac14\bigl(1+G_{12}+G_{13}+G_{23}\bigr).
\end{equation}
Each nonidentity element has traces $1,6,36,36,36,6,1$ in the same fermion-number sectors,
which sum to $\Tr_{\mathcal V}G_{ab}=122$. Hence
\begin{equation}
\dim\mathfrak F=\Tr_{\mathcal V}P_{\mathcal F}
=\frac{7898+3\times122}{4}=2066.
\end{equation}

\section{$O(r^{-6})$ correction to the tail}\label{r_6_tail}
We continue the change of variables in Section~\ref{tail} to obtain the $r^{-6}$ term in~\eqref{subleading_tail}. Since $P(x)=Ax^{-10}[1+O(x^{-9})]$, we need the relation between $x$ and $r=\rt\,s_1$ through order $x^{-5}$, including the fluctuations of $r$ at fixed $x$. 

Recall that $r$ is the largest eigenvalue separation of $\hat n^IX^I$ over unit vectors $\hat n$. From~\eqref{eq:param},
\begin{equation*}
r^2=\max_{\hat n^2=1}\Big[(\hat n\cdot\vec x)^2+2(\hat n\cdot u)^2+2(\hat n\cdot v)^2\Big].
\end{equation*}
Write $u_\parallel=\hat x\cdot u$ and $u_\perp=u-u_\parallel\hat x$, and similarly for $v$. Setting $\hat n=(1-\varepsilon^2/2)\hat x+\varepsilon+\dots$, with $\hat x\cdot\varepsilon=0$, the terms depending on $\varepsilon$ are $-x^2\varepsilon^2+4\varepsilon\cdot w$, where $w=u_\parallel u_\perp+v_\parallel v_\perp$. The maximum is at $\varepsilon=2w/x^2$, giving
\begin{equation}\label{eq:r2next}
r^2=x^2+2u_\parallel^2+2v_\parallel^2+\frac{4w^2}{x^2}+\dots.
\end{equation}
As in Section~\ref{tail}, the leading oscillator averages are
\begin{equation*}
\evs{u_Iu_J}_x=\evs{v_Iv_J}_x=\frac{\delta_{IJ}}{2x},\qquad
\evs{w^2}_x=\frac4{x^2},\qquad
\evs{(u_\parallel^2+v_\parallel^2)^2}_x=\frac2{x^2}.
\end{equation*}
Here the subscript $x$ means that the diagonal separation is held fixed. The factor 4 comes from the eight directions perpendicular to $\hat x$. These averages suffice for the terms with four powers of $u,v$. For the quadratic terms, we also need the first correction to $\evs{u_\parallel^2}_x$ and $\evs{v_\parallel^2}_x$.

We use the same background field gauge as in Section~\ref{tail}. The off-diagonal gauge field $A$ and the gauge fixing function $G$ are
\begin{equation*}
A=\begin{pmatrix}0&(a+i b)/\rt\\ (a-i b)/\rt&0\end{pmatrix},\qquad
G=\partial_\tau A-i[B^I,X^I],\qquad B^I=\frac{x^I\sigma_3}{2}.
\end{equation*}
Adding $\Tr G^2/2$ to the Lagrangian gives
\begin{equation}\label{eq:LB}
\begin{aligned}
L_B={}&\frac{\dot{\vec x}^{\,2}}4+2\dot{\vec x}\cdot(bu-av)+(bu-av)^2
+\frac12\big(\dot u^2+\dot v^2+\dot a^2+\dot b^2\big)\\
&+\frac{x^2}{2}\big(u^2+v^2+a^2+b^2\big)
+u^2v^2-(u\cdot v)^2.
\end{aligned}
\end{equation}
Thus all nine components of $u,v$, and $a,b$, oscillate with frequency $x$. Completing the square in $\dot{\vec x}$ gives $\tfrac14(\dot{\vec x}+4(bu-av))^2-3(bu-av)^2$. The remaining quartic terms are $u^2v^2-(u\cdot v)^2-3(bu-av)^2$. Taking the ground-state averages of $v,a,b$ gives the quadratic term in $u$,
\begin{equation*}
u^2\evs{v^2}_x-u_Iu_J\evs{v_Iv_J}_x-3u^2\evs{b^2}_x=\frac{5}{2x}\,u^2.
\end{equation*}
The gauge fixing also gives the Faddeev--Popov determinant,
\begin{equation*}
\det(-\partial_\tau^2+x^2-2u^2)\,\det(-\partial_\tau^2+x^2-2v^2).
\end{equation*}
Expanding $-\ln\det$ to first order in $u^2,v^2$, with $(-\partial_\tau^2+x^2)^{-1}(\tau,\tau)=1/(2x)$, adds $(u^2+v^2)/x$ to $L_B$. Together these terms change the frequency of a component $y$ of either $u$ or $v$ to
\begin{equation*}
\omega^2=x^2+\frac7x,\qquad
\evs{y^2}=\frac1{2\omega}=\frac1{2x}-\frac7{4x^4}+\dots.
\end{equation*}
We next include the Yukawa interaction. In the parametrization~\eqref{eq:param}, the fermion Hamiltonian is
\begin{equation}\label{eq:HF}
F=-x\,\chi^\dagger(\hat x\cdot\gamma)\chi
+(u^I-i v^I)\theta\gamma^I\chi
-(u^I+i v^I)\theta\gamma^I\chi^\dagger.
\end{equation}
For $y=u^I$, with $I$ fixed, the interaction is $y\,\theta\gamma^I(\chi-\chi^\dagger)$. The factor $y$ takes the bosonic ground state to the first excited state, adding energy $\omega$. The fermion operator fills an empty level or empties a filled one, adding energy $x$. Thus $E_n-E_0=\omega+x$, and second-order perturbation theory gives
\begin{equation*}
\delta E(\omega)=-\sum_n\frac{\big|\bra n\,y\,\theta\gamma^I(\chi-\chi^\dagger)\ket{0_{\hat x}}\big|^2}{E_n-E_0}
=-\frac1{2\omega}\,\frac8{\omega+x}.
\end{equation*}
The two factors in the numerator are $\evs{y^2}=1/(2\omega)$ and $\theta_\alpha\theta_\alpha=8$. Since the oscillator Hamiltonian contains $\omega^2y^2/2$, differentiating the energy at fixed $x$ gives $dE/d\omega=\omega\evs{y^2}$. The Yukawa interaction therefore changes the average by
\begin{equation*}
\delta\evs{y^2}=\left.\frac1\omega\frac{d\,\delta E}{d\omega}\right|_{\omega=x}=\frac3{x^4}.
\end{equation*}
The same calculation applies to $v$.

Finally, the derivatives in $H_0=-\nabla_{\vec x}^2$ also act on the oscillator wavefunction. For the coordinate $y$, the leading Hamiltonian is $h_y=-\tfrac12\partial_y^2+\tfrac12x^2y^2$, with ground state $\phi_0(y;x)=(x/\pi)^{1/4}e^{-xy^2/2}$ and energy $x/2$. Differentiating this Gaussian gives
\begin{equation*}
\partial_x\phi_0=\left(\frac1{4x}-\frac{y^2}{2}\right)\phi_0,\qquad
\partial_x^2\phi_0=\left(\frac{y^4}{4}-\frac{y^2}{4x}-\frac3{16x^2}\right)\phi_0.
\end{equation*}
To solve for the correction, we write these polynomials in terms of the unnormalized oscillator wavefunctions
\begin{align*}
\phi_2&=(2xy^2-1)\phi_0,&(h_y-x/2)\phi_2&=2x\phi_2,\\
\phi_4&=(4x^2y^4-12xy^2+3)\phi_0,&(h_y-x/2)\phi_4&=4x\phi_4.
\end{align*}
Their energies are $2x$ and $4x$ above that of $\phi_0$. Using $H_0\Psi_0=0$ and $\partial_x\Psi_0=-9\Psi_0/x$ from~\eqref{eq:tail},
\begin{equation*}
H_0(\Psi_0\phi_0)=-\Psi_0\left(\partial_x^2-\frac{10}{x}\partial_x\right)\phi_0
=\frac{\Psi_0}{x^2}\left(\frac18\phi_0-\frac{11}{4}\phi_2-\frac1{16}\phi_4\right).
\end{equation*}
Dividing by these energy differences, we cancel the $\phi_2$ and $\phi_4$ terms in $H\Psi=0$ by adding
\begin{equation*}
\delta\phi_0=\frac{11}{8x^3}\phi_2+\frac1{64x^3}\phi_4.
\end{equation*}
A correction proportional to $\phi_0$ cancels from the normalized average of $y^2$. Since $\phi_2,\phi_4$ are orthogonal to $\phi_0$, they do not change the normalization at this order. The Gaussian integrals give $\int dy\,y^2\phi_0\phi_2=1/x$ and $\int dy\,y^2\phi_0\phi_4=0$, so
\begin{equation*}
\delta\evs{y^2}=2\int dy\,y^2\phi_0\delta\phi_0=\frac{11}{4x^4}.
\end{equation*}
Adding the three corrections, we obtain
\begin{equation}\label{eq:spread8}
\evs{u_\parallel^2}_x=\evs{v_\parallel^2}_x
=\frac1{2x}-\frac7{4x^4}+\frac3{x^4}+\frac{11}{4x^4}
=\frac1{2x}\left(1+\frac8{x^3}\right)+\dots.
\end{equation}
Taking the square root of~\eqref{eq:r2next},
\begin{equation*}
r=x+\frac{u_\parallel^2+v_\parallel^2}{x}
+\frac{2w^2-\tfrac12(u_\parallel^2+v_\parallel^2)^2}{x^3}+\dots.
\end{equation*}
Using~\eqref{eq:spread8} in the quadratic term and the leading averages in the quartic terms gives, for $\delta r=r-x$,
\begin{equation*}
\evs{\delta r}_x=\frac1{x^2}+\frac{8+8-1}{x^5}+\dots
=\frac1{x^2}+\frac{15}{x^5}+\dots,\qquad
\evs{(\delta r)^2}_x=\frac2{x^4}+\dots.
\end{equation*}
The second moment includes the fluctuations about the mean, $\evs{(r-\evs{r}_x)^2}_x=1/x^4+\dots$. To obtain the probability per unit $r$, we average over the shift and expand the delta function to second order. Using $P(x)=Ax^{-10}$ to this order,
\begin{align*}
P(r)&=\int dx\,P(x)\,\evs{\delta(r-x-\delta r)}_x\\
&=\left[P(x)-\frac{d}{dx}\big(P(x)\evs{\delta r}_x\big)
+\frac12\frac{d^2}{dx^2}\big(P(x)\evs{(\delta r)^2}_x\big)\right]_{x=r}+\dots\\
&=\frac{A}{r^{10}}-A\frac{d}{dr}\left(\frac1{r^{12}}+\frac{15}{r^{15}}\right)
+A\frac{d^2}{dr^2}\frac1{r^{14}}+\dots\\
&=\frac{A}{r^{10}}\left(1+\frac{12}{r^3}+\frac{435}{r^6}+O(r^{-9})\right).
\end{align*}

\section{Herderschee--Maldacena amplitude}\label{app:HM}
In this appendix we obtain the value $g_{\rm 3pt}$ of $g$ in~\eqref{eq:Agq} from the three-point amplitude of~\cite{Herderschee:2023pza}. In \cite{Herderschee:2023pza}  the coordinates used are the physical positions and light-cone time $x^+$, with $x^\pm=(x^0\pm x^{10})/\rt$ and $x^-$ compact with radius $R_-$, so that $N$ D0-branes have $-p_-=N/R_-$. With the same fermions, our Hamiltonian and theirs are
\begin{equation}
\begin{split}
H&=\Tr\Big[\frac12P_I^2-\frac14[X^I,X^J]^2-\frac12\psi\gamma^I[X^I,\psi]\Big],\\
H_{\rm HM}&=R_-\Tr\Big[\frac12P_{{\rm HM},I}^{\,2}
-\frac{[X_{\rm HM}^I,X_{\rm HM}^J]^2}{4(2\pi)^2\ell_p^6}
-\frac{\psi\gamma^I[X_{\rm HM}^I,\psi]}{2(2\pi)\ell_p^3}\Big].
\end{split}
\end{equation}
Setting $X_{\rm HM}=LX$ and $P_{\rm HM}=P/L$ and matching the quartic term gives
\begin{equation}\label{eq:HMunits}
L=(2\pi)^{1/3}\ell_p,\qquad H_{\rm HM}=\frac{R_-}{L^2}H,
\end{equation}
so the momenta of~\cite{Herderschee:2023pza} are $\vec p/L$, with $\vec p$ our momentum.

The amplitude is nonzero only when $p_+=0$ and the transverse momenta are $\vec p_k=q_k\epsilon$, with $\epsilon$ as in Section~\ref{sec:3pt-LS}. Particle $k$ has the spinor
\begin{equation}
\lambda_k=\Big(-\sqrt{\frac{N_k}{R_-}},\ \frac{q_k}{L}\sqrt{\frac{R_-}{N_k}}\Big),\qquad
\langle kl\rangle=\lambda_k^1\lambda_l^2-\lambda_k^2\lambda_l^1=\frac{q_kN_l-q_lN_k}{L\sqrt{N_kN_l}},
\end{equation}
and the amplitude is
\begin{equation}
\mathcal A_3=(2\pi)^{11}\delta\Big(\sum_kp_{k-}\Big)\delta\Big(\sum_kp_{k+}\Big)\delta^9\Big(\sum_k\frac{\vec p_k}{L}\Big)\,M_3,\qquad
M_3=(2\pi)^4\ell_p^{9/2}\,\rt\,\frac{\delta^{16}\big(\sum_k\lambda_k\eta_k\big)}{\big(\langle12\rangle\langle23\rangle\langle13\rangle\big)^2},
\end{equation}
where the Grassmann variables $\eta_{km}$, $m=1,\dots,8$, label the polarization of particle $k$.

For two D0-branes joining into the bound state,
\begin{equation}
N_1=N_2=1,\qquad N_3=2,\qquad q_1=-q_2=q,\qquad q_3=0,
\end{equation}
the brackets are
\begin{equation}
\langle12\rangle=\frac{2q}{L},\qquad\langle23\rangle=-\frac{\rt q}{L},\qquad\langle13\rangle=\frac{\rt q}{L}.
\end{equation}
With $\eta^{\rm cm}=(\eta_1+\eta_2)/\rt$ and $\eta^{\rm rel}=(\eta_1-\eta_2)/\rt$, the two components of the argument of the delta function are
\begin{equation}
\sum_k\lambda_k^1\eta_{km}=-\sqrt{\frac{2}{R_-}}\big(\eta^{\rm cm}_m+\eta_{3m}\big),\qquad
\sum_k\lambda_k^2\eta_{km}=\sqrt{2R_-}\,\frac qL\,\eta^{\rm rel}_m.
\end{equation}
The Grassmann delta function and the stripped amplitude become
\begin{equation}\label{eq:M3}
\begin{split}
\delta^{16}\Big(\sum_k\lambda_k\eta_k\Big)&=256\,\frac{q^8}{L^8}\,\delta^8(\eta^{\rm cm}+\eta_3)\,\delta^8(\eta^{\rm rel}),\\
M_3&=16\rt\,(2\pi)^4\ell_p^{9/2}\,\frac{q^2}{L^2}\,\delta^8(\eta^{\rm cm}+\eta_3)\,\delta^8(\eta^{\rm rel}).
\end{split}
\end{equation}
The first delta function matches the centre of mass polarization of the two D0-branes to that of the bound state. The second selects the relative polarization and gives the factor $\braket{-}{a}$ in~\eqref{eq:Agq}.

We now write $\mathcal A_3$ in our variables and in the normalization of our states. With $x^-$ compact, $\delta(\sum_kp_{k-})=R_-$, and by~\eqref{eq:HMunits} the energies of~\cite{Herderschee:2023pza} are $R_-/L^2$ times ours, so $\delta(\sum_kp_{k+})=(L^2/R_-)\,\delta(\Delta E)$. Also $\delta^9(\sum_k\vec p_k/L)=L^9\delta^9(\Delta\vec P)$. The delta functions in $\mathcal A_3$ are therefore
\begin{equation}
2\pi L^{11}\cdot2\pi\,\delta(\Delta E)\,(2\pi)^9\delta^9(\Delta\vec P),
\end{equation}
 The one-particle states of~\cite{Herderschee:2023pza} are normalized to
\begin{equation}
-2p_-(2\pi)^{10}\delta(p_--p_-')\,\delta^9\big((\vec p-\vec p\,')/L\big)=4\pi N\,L^9\,\delta_{NN'}\,(2\pi)^9\delta^9(\vec p-\vec p\,'),
\end{equation}
while our states, with wavefunction $e^{\i\vec p\cdot\vec r}$, are normalized to $\braket{\vec p\,'}{\vec p}=(2\pi)^9\delta^9(\vec p-\vec p\,')$.

So each state in~\cite{Herderschee:2023pza} is related to ours by a factor $\sqrt{4\pi N}\,L^{9/2}$. Dividing by this factor for $N=1,1,2$ and using~\eqref{eq:M3},
\begin{equation}
\mathcal A=\frac{2\pi L^{11}}{\sqrt{4\pi}\sqrt{4\pi}\sqrt{8\pi}\,L^{27/2}}\,M_3=\frac{2\pi}{\sqrt{4\pi}\sqrt{4\pi}\sqrt{8\pi}}\,16\rt\,(2\pi)^4\,\frac{\ell_p^{9/2}}{L^{9/2}}\,q^2\,\braket{-}{a}.
\end{equation}
This is the amplitude for the normalized gauge-invariant state $\ket{\vec p,a}_{\rm sym}$ of~\eqref{eq:exchange}. That state is $\rt$ times the projection of $\ket{\vec p,a}$ onto gauge-invariant states, and $\ket\Omega$ is gauge invariant, so $\mathcal A=\rt\,\mathcal A_\text{3pt}$ with $\mathcal A_\text{3pt}$ of~\eqref{eq:Agq}. Using $L^{9/2}=(2\pi)^{3/2}\ell_p^{9/2}$,
\begin{equation}
g_{\rm 3pt}=16\pi^2,
\end{equation}
and~\eqref{eq:tailr} becomes
\begin{equation}
\Psi(\vec r)\simeq\frac{945}{2\pi^2}\,\frac{\hat r_i\hat r_j}{r^9}\ket{ij},\qquad A_{\rm 3pt}=\frac{945\,g_{\rm 3pt}^2}{4\pi^4}=60480.
\end{equation}

\section{Details of the wavefunction numerics}\label{app:ground-numerics}
This appendix gives implementation details for the
numerical solution of the ground state in Section~\ref{ground_solve}.
We first express the supercharges in singular-value
coordinates and describe the boundary conditions required
for a smooth wavefunction. We then specify the numerical
grid, polynomial interpolation, and iterative procedure
used to minimize $\mathcal F/\mathcal{N}$. %
We also visualize the probability distribution in other ways and compare them with the Born-Oppenheimer approximation along the valleys. 
Finally, we discuss some additional details of the algorithm at finite energies and non-singlet irreps of SO(9).

\subsection{Angular derivatives and boundary conditions}
To write $Q_\alpha\Omega=0$ in this gauge, we have to express all the derivatives $P_a^I=-i\partial/\partial X_a^I$ using only $s_1,s_2,s_3$. At $X_a^I=s_a\delta_a^I$, the three diagonal derivatives are
\[
\frac{\partial\Omega}{\partial X_a^a}=\frac{\partial\Omega}{\partial s_a}\qquad\text{(no sum)}.
\]
For distinct nonzero $s_a$, the other 24 derivatives change the orientation of $X$. They are not zero, the fermionic wavefunction changes with that orientation, even though the full state is a singlet. The bosonic and fermionic rotation generators therefore sum to zero. SO(9) and SU(2) invariance give, for $\mu=4,\dots,9$ and $a\ne b$,
\[
\begin{aligned}
\left(s_a\frac{\partial}{\partial X_a^\mu}+\Sigma_{a\mu}\right)\Omega&=0&&\mathrm{SO}(9),\\
\left(s_a\frac{\partial}{\partial X_a^b}-s_b\frac{\partial}{\partial X_b^a}+\Sigma_{ab}\right)\Omega&=0&&\mathrm{SO}(9),\\
\left(s_a\frac{\partial}{\partial X_b^a}-s_b\frac{\partial}{\partial X_a^b}+\Lambda_{ab}\right)\Omega&=0&&\mathrm{SU}(2).
\end{aligned}
\]
Solving for the angular derivatives gives
\begin{equation}
\begin{split}
\frac{\partial\Omega}{\partial X_a^\mu}
&=-\frac{\Sigma_{a\mu}}{s_a}\Omega,\qquad \mu=4,\dots,9,\\
\frac{\partial\Omega}{\partial X_a^b}
&=-\frac{s_a\Sigma_{ab}+s_b\Lambda_{ab}}{s_a^2-s_b^2}\Omega,\qquad a\ne b.
\end{split}
\end{equation}

At $s_a=s_b>0$, the last two singlet equations become dependent and require $(\Sigma_{ab}+\Lambda_{ab})\Omega=0$ on the wall. At $s_a=0$, the first equation requires $\Sigma_{a\mu}\Omega=0$. These are necessary boundary conditions at the walls since the wavefunction must remain smooth in the original coordinates $X_a^I$.

\subsection{Grid and interpolation}

We take a separate ansatz for  $\chi_t$ in the ``core'' $s_1 < S$ and ``tail'' region $s_1> S$. We chose $S = 9/2^{2/3}$. 
We also introduce the parameter $S_{\max} =60/2^{2/3}$.
In each region $e$, we use $-1\leq x\leq 1$, with
\[
s_1^{(\mathrm{core})}(x)=\frac{S(1+x)}{3-3x+2},
\qquad
s_1^{(\mathrm{tail})}(x)=S\left(\frac{S_{\max}}{S}\right)^{(1+x)/2}.
\]
The other two variables are the polar angle
$\tan \theta=s_3/s_2$ and 
$u=\sqrt{s_2^2+s_3^2}/\rho_{\max}(s_1,\theta)$, where $\rho_{\max}$
is the transverse cutoff.

We choose $\chi_t$ to be a polynomial interpolation on a discrete set of fixed sample points; the value of $\chi_t$ on each point is what enters the variational method. We choose the sample points separately in the core and the tail.

More explicitly, in each region $e$, the points $x_1,\ldots,x_{N_e}$ are the zeros
of the Legendre polynomial $P_{N_e}$ on $[-1,1]$. For $u$ and
$\theta$, we take the zeros of $P_{N_u}$ and $P_{N_\theta}$ and
rescale them to $[0,1]$ and $[0,\pi/4]$, respectively.
Within each region, we write
\begin{equation}\label{eq:polynomialInterpolation}
\chi_t(s_1,s_2,s_3) =
\sum_{i=1}^{N_e}\sum_{j=1}^{N_u}\sum_{k=1}^{N_\theta}
\chi_{t,ijk}^{(e)}
\prod_{m\ne i}\frac{x-x_m}{x_i-x_m}
\prod_{n\ne j}\frac{u-u_n}{u_j-u_n}
\prod_{p\ne k}\frac{\theta-\theta_p}{\theta_k-\theta_p}.
\end{equation}
At the sample point $(x_i,u_j,\theta_k)$, the product multiplying
$\chi_{t,ijk}^{(e)}$ is one and all other terms vanish. Thus the
quantities that enter the variational method are simply the values of $\chi_t$ at the sample
points, which are $\chi_{t,ijk}$.

We use $N_{\rm core}=276$, $N_{\rm tail}=40$,
$N_u=80$, and $N_\theta=32$.
There are thus $(276+40)\times 80 \times32=808\,960$ nodes and 2066 components at each node, giving $\approx 1.7\times10^9$ variational parameters.

\subsection{Iterative solution}
In each cycle we solve the reduced eigenvalue problem by an iterative method for its lowest eigenvector (the so-called LOBPCG\footnote{This stands for Locally Optimal Block Preconditioned Conjugate Gradient \url{https://en.wikipedia.org/wiki/LOBPCG}. } algorithm). The fermion basis contains the 200 combinations carrying the most norm, together with combinations obtained from the derivatives of $\mathcal{F}/\mathcal{N}$ outside this space. We add 30 at a time, up to 230 in total. Each cycle has 800 iterations, after which the basis is rebuilt. We evaluate ${\mathcal F}$ using all 2066 components as a check on the reduced calculation. One round consists of two such cycles. The final state is obtained from three rounds followed by five updates of the outer region.

Let $\mathbf c_1$ and $\mathbf c_2$ denote the grid values in the core and outer region, respectively, and write $A$ in the corresponding two-by-two block form. With $\mathbf c_1$ held fixed, minimizing ${\mathcal F}$ over $\mathbf c_2$ gives
\begin{equation}
A_{22}\mathbf c_2=-A_{21}\mathbf c_1.
\end{equation}
Only the term joining the two regions contributes to $A_{21}$. We solve this linear system in all 2066 fermion components. 
\subsection{More properties of the numerical solution}

\begin{figure}[!htb]
\centering
\includegraphics[width=\textwidth]{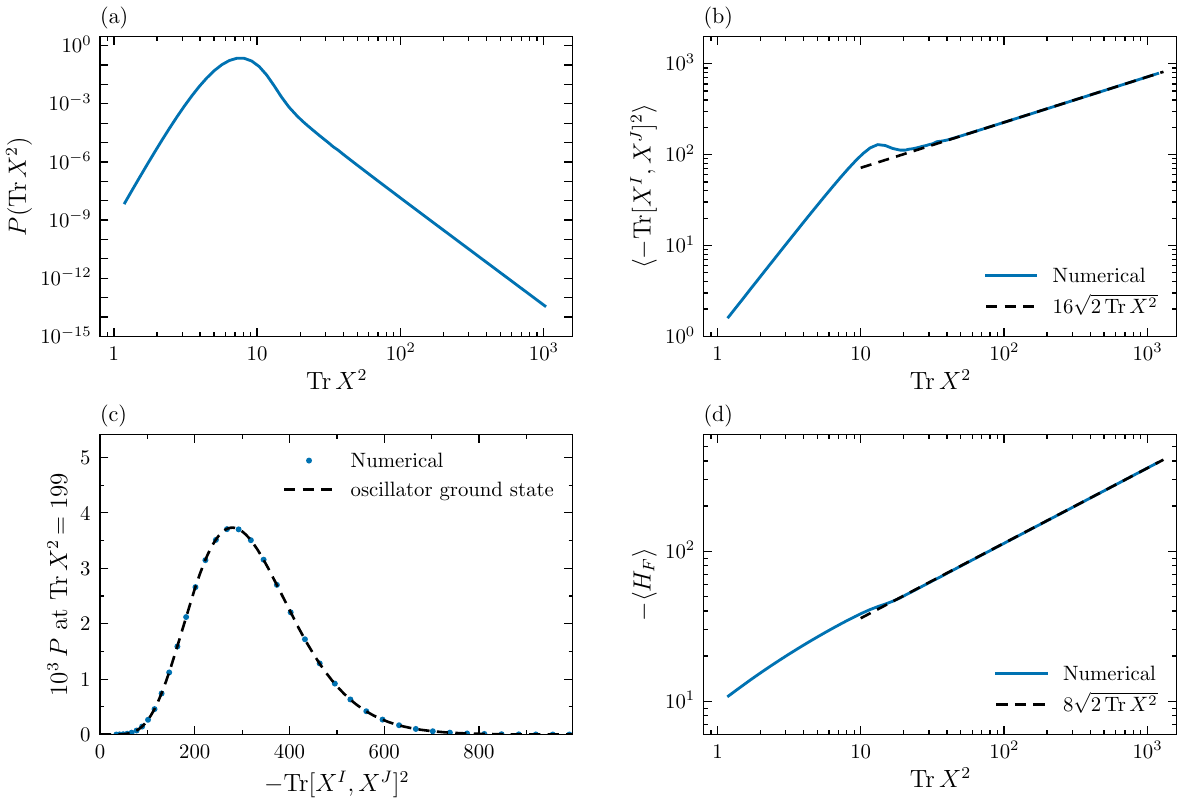}
\caption{The numerical ground state. (a) Probability density of $t=\Tr X^2$. (b), (d) Averages of $-\Tr[X^I,X^J]^2$ (summed over $I,J$) and $-F$ at fixed $t$. (c) Distribution of $-\Tr[X^I,X^J]^2$ at $r\simeq20$, for which $t\simeq199$ along the valley. Dashed curves show the oscillator predictions derived in the text.}
\label{fig:wavefunction}
\end{figure}

    To describe configurations of different sizes, we use $t=\Tr X^2$. Figure~\ref{fig:wavefunction} shows its probability density $P(t)$ and averages of operators at fixed $t$, defined by
\begin{equation}
\begin{aligned}
P(t)&=\int d\mu\,\Omega^\dagger\Omega\,\delta\big(\Tr X^2-t\big),\\
\evs{O}_t&=\frac{1}{P(t)}\int d\mu\,\Omega^\dagger O\,\Omega\,\delta\big(\Tr X^2-t\big).
\end{aligned}
\end{equation}
 Panel (a) shows this distribution, while panels (b) and (d) show $\evs{-\Tr[X^I,X^J]^2}_t$ and $-\evs{F}_t$. The large-$t$ behavior follows from the oscillator ground state along the valleys. Here $s_1\gg s_2,s_3$ and the brane separation is $r\simeq\sqrt{2t}$. There are sixteen real off-diagonal bosonic coordinates $y_k$. To leading order they are oscillators of frequency $r$, whose probability density at fixed $r$ and potential are
\begin{equation}
p(\mathbf y\mid r)=\left(\frac{r}{\pi}\right)^8e^{-r y^2},\qquad
V\simeq\frac{r^2y^2}{2}\,
\end{equation}
 Each oscillator has $\evs{y_k^2}=1/(2r)$, so $\evs{V}\simeq4r$. The sixteen complex fermionic partners have ground-state energy $-8r$, which cancels the bosonic zero-point energy $16\times r/2$. Hence
\begin{equation*}
\evs{-\Tr[X^I,X^J]^2}_t\simeq16\sqrt{2t},\qquad
-\evs{F}_t\simeq8\sqrt{2t}.
\end{equation*}
These give the dashed curves in panels (b) and (d). Panel (c) tests the full distribution of $-\Tr[X^I,X^J]^2=4V$ at $r\simeq20$. 

In the oscillator approximation, the normalized density of $\zeta=4V/r=2r\,y^2$, with $y^2=y_ky_k$, is
\begin{equation*}
p(\zeta)=\frac{\zeta^7e^{-\zeta/2}}{2^8\,7!},\qquad \zeta\ge0.
\end{equation*}
The power $\zeta^7$ comes from the radial measure in sixteen dimensions. The density plotted for $4V$ is therefore $p(4V/r)/r$. At smaller $t$, where the three singular values are comparable, the separation into relative motion and off-diagonal oscillators no longer applies; the numerical solution includes this region as well.
\begin{figure}[htbp]
\centering
\includegraphics[width=\textwidth]{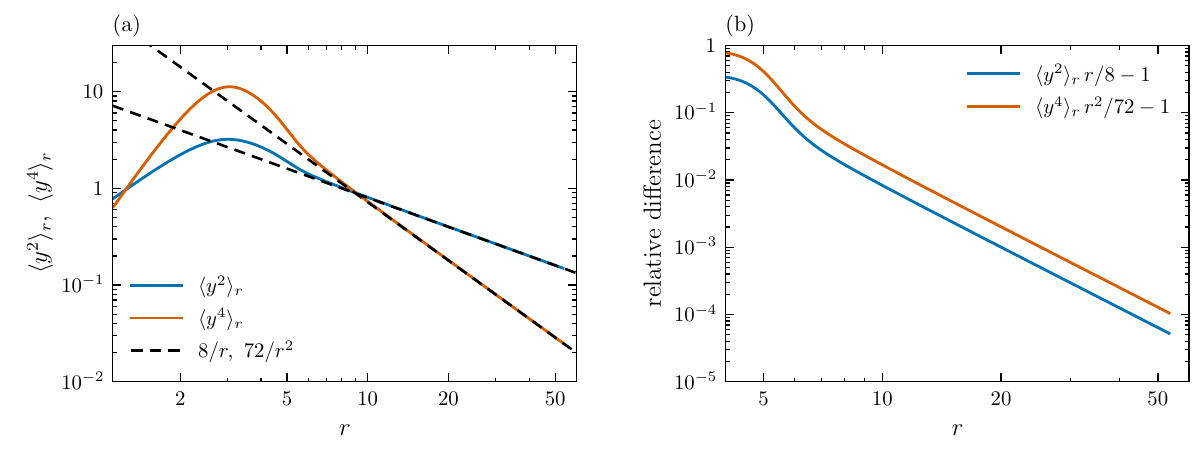}
\caption{(a) The averages $\evs{y^2}_r$ and $\evs{y^4}_r$ at fixed $r$, where $y^2=s_2^2+s_3^2$, with their values $8/r$ and $72/r^2$ for the off-diagonal oscillators in their ground state (dashed). (b) The relative differences $\evs{y^2}_r\,r/8-1$ and $\evs{y^4}_r\,r^2/72-1$.}
\label{fig:moments}
\end{figure}

\subsection{Positive-energy states}
\begin{table}
\centering
\begin{tabular}{crr}
\toprule
SO(9) irrep & Highest weight & dim $\mathfrak{F}_J$ \\
\midrule
$\mathbf{1}$     & $(0,0,0,0)$  & $2\,066$ \\
$\mathbf{9}$     & $(1,0,0,0)$  & $9\,834$ \\
$\mathbf{36}$    & $(1,1,0,0)$  & $23\,352$ \\
$\mathbf{44}$    & $(2,0,0,0)$  & $27\,540$ \\
$\mathbf{450}$   & $(4,0,0,0)$  & $92\,490$ \\
$\mathbf{495}$   & $(2,2,0,0)$  & $93\,567$ \\
$\mathbf{2508}$  & $(6,0,0,0)$  & $198\,332$ \\
$\mathbf{4004}$  & $(3,3,0,0)$  & $217\,594$ \\
$\mathbf{9867}$  & $(8,0,0,0)$  & $345\,072$ \\
$\mathbf{30888}$ & $(10,0,0,0)$ & $532\,710$ \\
\bottomrule
\end{tabular}
\caption{Number of independent wavefunction components
at each point $(s_1,s_2,s_3)$ for the total SO(9)
representations used in our calculations.
These numbers do not include the SO(9) multiplet degeneracy factors.}
\label{tab:sector-components}
\end{table}
We compute positive-energy states using the same
singular-value coordinates and polynomial interpolation,
but minimize the expectation value of $H$ rather than
the weighted functional \eqref{eq:ground-least-squares}.
For each total SO(9) representation $J$, we impose
the hard-wall boundary condition of
\eqref{eq:smatrix-box-schrodinger} and compute several
of the lowest levels together. The variational functional is
\begin{equation}
\mathcal E[\Psi]
=\frac{\bra{\Psi}H\ket{\Psi}}{\braket{\Psi}{\Psi}}.
\end{equation}
For a nontrivial total SO(9) representation $J$, we may still
use the three singular values, but rotations also act
on the state's $j$ index $\ket{E,J,j}$, where $j$ takes $\dim J$ values. 
Then the residual gauge symmetry discussed around \eqref{residualGauge} now requires that the fermions transform covariantly under these SO(9) generators (as opposed to be invariant). Let $t^{(J)}_{KL}$ denote the anti-Hermitian SO(9)
generators , they act on the $j$ indices
$t^{(J)}_{KL} \ket{E,J,j}=(t^{(J)}_{KL})_{j'j} \ket{E,J,j'}$.
The residual symmetry constraints \eqref{eq:singlet-stabilizer} are now modified to
\begin{equation}
\left(\Sigma_{KL}+t^{(J)}_{KL}\right)\Psi_J=0
\quad(K,L=4,\dots,9),\qquad
e^{\pi(\Sigma_{ab}+\Lambda_{ab}+t^{(J)}_{ab})}
\Psi_J=\Psi_J
\quad(1\leq a<b\leq3).
\end{equation}
For the singlet, $t^{(J)}=0$, recovering \eqref{eq:singlet-stabilizer}
that define the 2066-dimensional space $\mathfrak F$.
For nontrivial $J$, the combined constraints give the
component counts $\dim \mathfrak{F}_J$ in Table~\ref{tab:sector-components}.

Another subtlety that occurs at finite energy is the following.
To find an excited level, we must require the
state to be orthogonal to the lower states. Otherwise,
minimization would return the lowest state again.
In practice, this can be accomplished by the ``block'' LOBPCG algorithm, where one computes several levels together. 
This algorithm maintains the mutual
orthogonality of the trial eigenvectors with respect to the discretized inner product
$\braket{\Psi_m}{\Psi_n}=\mathbf c_m^\dagger M\mathbf c_n$.

To keep the calculation manageable, we perform each
block LOBPCG solve in a reduced subspace.
Between cycles, we rebuild this basis from the fermion
combinations carrying most of the norm of the current
states. We also add combinations obtained from their
eigenvector residuals,
\begin{equation}
\ket{r_n}=(H-E_n)\ket{\Psi_n}.
\end{equation}
We select additional fermion combinations from their
components outside the current basis.
We then repeat the block LOBPCG solve in the updated
basis and monitor convergence. This basis update is an additional step beyond the standard LOBPCG algorithm.%
The resulting spectrum includes both scattering levels
and localized states, which we analyze using the tests
described in Section~\ref{sec:numerical}.

\section{Finite-box corrections to the scattering phases}\label{app:smatrix-finite-box}

We first account for the finite position of the wall in the relations of Section~\ref{sec:smatrix-box}. There are two corrections to the radius used in the free radial solution, followed by a correction for the interaction beyond the wall. The first follows from $r=x+x^{-2}+\dots$ in Section~\ref{tail}. A wall at $r=R_c$ is therefore at
\begin{equation}\label{eq:smatrix-radial-shift}
x(R_c)=R_c-\frac1{R_c^2}+\dots
\end{equation}
in the coordinate $x$ of the free radial motion.

The second correction arises because the wall requires the entire wavefunction to vanish, including its excited oscillator components. We use the ground state $\ket{0_{\hat r}}$ of Section~\ref{sec:3pt-LS}, whose sixteen transverse oscillators have frequency $r=\rt\,s_1$. Appendix~\ref{r_6_tail} uses background field gauge, where the leading frequency is $x$. The bosonic part of $\ket{0_{\hat r}}$ is the product of the sixteen Gaussians $\phi_0$ given there, with $x$ replaced by $r$. Differentiating at fixed $\hat r$ gives
\begin{equation*}
\partial_r\ket{0_{\hat r}}=\left(\frac4r-\frac{u^2+v^2}{2}\right)\ket{0_{\hat r}},\qquad
\big\|\partial_r\ket{0_{\hat r}}\big\|^2=\frac{16}{8r^2}=\frac2{r^2}.
\end{equation*}
Here $u,v$ are the off-diagonal coordinates, each with eight components perpendicular to $\hat r$, as in~\eqref{eq:gauge3pt}. As in Appendix~\ref{r_6_tail}, the derivative is a sum of states with one oscillator in its second excited state and the others in their ground state. Each term is $2r$ above the ground state. Denote the normalized combination by $\ket{2_{\hat r}}=-r\,\partial_r\ket{0_{\hat r}}/\sqrt2$.

Write these two components of the radial wave as $f(r)\ket{0_{\hat r}}+\tilde f(r)\ket{2_{\hat r}}$, using the radial normalization of~\eqref{eq:radial}, so their contribution to the probability is $(|f|^2+|\tilde f|^2)\,dr$. When the radial kinetic energy acts on this expression, one derivative on a coefficient and one on an oscillator state gives
\begin{equation}\label{eq:smatrix-oscillator-pair}
-f''-\frac{2\sqrt2}{r}\tilde f'=Ef,\qquad
-\tilde f''+\frac{2\sqrt2}{r}f'+2r\tilde f=E\tilde f.
\end{equation}
We have kept the terms needed for the leading wall correction. The numerical solve uses the full Hamiltonian, including the centrifugal terms and the other oscillator components.

Away from the wall, $f''\simeq-Ef$, and the second equation gives $\tilde f\simeq-\sqrt2\,f'/r^2$. Thus setting $f$ to zero does not also set $\tilde f$ to zero. Near the wall the excited component changes on the short length $(2R_c-E)^{-1/2}$. For $E\ll2R_c$ we can hold the coefficients fixed at $r=R_c$. To cancel the value $\tilde f(R_c)\simeq-\sqrt2\,f'(R_c)/R_c^2$ continued from the interior, we add
\begin{align*}
\delta\tilde f(r)&\simeq\frac{\sqrt2}{R_c^2}f'(R_c)e^{\sqrt{2R_c-E}(r-R_c)},\\
\delta f(r)&\simeq-\frac{2\sqrt{2R_c-E}}{R_c^4}f'(R_c)e^{\sqrt{2R_c-E}(r-R_c)}.
\end{align*}
The second expression follows from the first equation in~\eqref{eq:smatrix-oscillator-pair}. Requiring $f(R_c)+\delta f(R_c)=0$ shifts the zero of the continued radial wave inward by $2\sqrt{2R_c-E}/R_c^4$. This gives the leading dependence on radius and energy. In the fit we allow its coefficient to vary,
\begin{equation}\label{eq:smatrix-effective-wall}
R_{{\rm eff},i}(E)=R_c-\frac1{R_c^2}-2c_i^{\rm wall}\frac{\sqrt{2R_c-E}}{R_c^4}.
\end{equation}
There is one fitted constant $c_i^{\rm wall}$ for each orbital component of a fixed $J$, shared by all energies and box sizes. The index $i$ labels the rows of $D_J$ in~\eqref{eq:smatrix-coupled-box}.

The box also omits the interaction at larger separations. For a $p^4/r^7$ term, the first-order singlet phase is proportional to $p^{-1}\int dr\,(p^4/r^7)[pr\,j_5(pr)]^2$. Setting $t=pr$ gives $p^9\int dt\,j_5(t)^2/t^5$. We subtract the part beyond the wall by using
\begin{equation}\label{eq:smatrix-tail-cut}
\delta_2^{\rm box}(E,R_c)=\delta_2(E)-75600\,a_2p^9\int_{pR_{\rm eff}}^\infty\frac{j_5(t)^2}{t^5}\,dt.
\end{equation}
The factor 75600 normalizes the integral from zero to infinity to one. Here $a_2$ is the fitted coefficient of $p^9$ in $\delta_2$; its value is not fixed to the supergravity prediction.\footnote{The $r^{-7}$ form and the use of the total phase coefficient $a_2$ to normalize this correction are assumptions of the fit. A threshold bound state also contributes at order $p^9$, so $a_2$ is not in general the coefficient of the long-range force alone.} We use the same correction for the independent $(2,2,0,0)$ sector, which also measures $\delta_2$.

For the box calculation, we use polynomial interpolation as in~\eqref{eq:polynomialInterpolation}, with $16\times12$ points in the two transverse coordinates and 48--88 radial points, depending on the box size. The grid values are restricted so that the interpolating polynomial vanishes at the outer wall. We evaluate the integrals on finer Gauss grids, with 2.25 times as many radial and angular points and nine times as many transverse radial points. The extra transverse points resolve the sharply concentrated oscillator probability. We solve for several levels together, updating the reduced fermion basis from the current states and the additional combinations in $H\Psi_n-E_n\Psi_n$.

We estimate the remaining iteration error by extrapolating the successive changes in each energy as a geometric sequence, and check the extrapolation by omitting the last iteration cycle. The errors $\sigma_n$ also include the change between earlier and later runs and the remaining grid error measured with finer grids. We repeat selected boxes with finer integration grids and larger fermion bases to check convergence.

We determine the phases by fitting all box sizes together. Near threshold we use a series with free coefficients,
\begin{equation*}
\delta_\ell(p)=a_\ell p^9+b_\ell p^{18}+c_\ell p^{27}+\dots.
\end{equation*}
These powers are motivated by the low-energy expansion in Section~\ref{sec:smatrix}. We also repeat the fit with the leading power free to test the $p^9$ behavior. For the first resonances in $\delta_0$ and $\delta_2$, we add
\begin{equation*}
\arg\!\left[E_r-E+\frac{\i\Gamma_r}{2}\left(\frac{E}{E_r}\right)^{15/2}\right],
\end{equation*}
with $E_r$ and $\Gamma_r$ fitted, following the phase continuously through $\pi/2$. The power $15/2$ gives the threshold behavior for $\ell=0$. For $\ell=2$ it is a choice of fitting function, whose effect we check by changing the power. At higher energies we continue the threshold series by a quadratic polynomial in $p$, matching its value and first two derivatives. This avoids the rapid growth of the high powers outside the range where they were fitted. We add piecewise cubic polynomials, joined with continuous first and second derivatives, to describe the remaining energy dependence.

To evaluate the box condition without poles from the Bessel ratios or tangents, we multiply them out in~\eqref{eq:smatrix-coupled-box} and use
\begin{equation*}
\det_{ij}\!\left[(V_J)_{ij}\left\{j_{\ell_i+3}(pR_{{\rm eff},i})\cos\delta_{\ell_j}-y_{\ell_i+3}(pR_{{\rm eff},i})\sin\delta_{\ell_j}\right\}\right]=0.
\end{equation*}
The row labels the orbital component and the column labels the phase. For the singlet this reduces to $j_5\cos\delta_2^{\rm box}-y_5\sin\delta_2^{\rm box}=0$. At a measured energy $E_n$, the value of the left side divided by its energy derivative gives minus the first-order energy change $\Delta E_n$ needed to satisfy the condition. We fit the phase parameters and wall coefficients by minimizing
\begin{equation*}
\chi^2=\sum_n\left(\frac{\Delta E_n}{\sigma_n}\right)^2.
\end{equation*}
Thus every term compares the predicted and computed energy in units of its numerical error. For the piecewise polynomial fits, we add a sum of squared second differences of the coefficients, meaning differences between successive coefficient changes. This discourages rapid variation where the levels leave the phase poorly constrained. If a box has more scatter than its quoted errors allow, we multiply its errors by $\max(1,\sqrt{\chi^2_{\rm box}/n_{\rm box}})$ and repeat the fit. Here $n_{\rm box}$ is the number of fitted levels in that box.

The threshold fit uses 304 levels and gives $\chi^2=210.7$ for 257 degrees of freedom. Including the $(2,2,0,0)$ boxes gives $\chi^2=220.4$ for 268 degrees of freedom, with every added level within 1.5 times its numerical error. The phases $\delta_2$ extracted separately from these boxes agree with the singlet curve to a rms difference of $10^{-4}$ radians below $E=6.5$. We repeat the fits with different powers, energy ranges, polynomial spacing and smoothness penalties, and with individual boxes removed.

\section{Details on the singlet bound state existence proof \label{app:boundState}}
On the positive-energy subspace define
\begin{equation}
C_\alpha=Q_\alpha H^{-1/2},\qquad
\{C_\alpha,C_\beta\}=2\delta_{\alpha\beta}.
\end{equation}
Using the multiplet decomposition \eqref{eq:supermultiplet}, we may define a projector $\Pi_{\mathbf{84}}$ onto the $\mathbf{84}$ component of $S$. This projector is a polynomial in $C_\alpha$ (and is defined to be 0 on the $H=0$ ground state.)
$\Pi_{\mathbf{84}}$ commutes with $H$ and rotations; on total SO(9) singlets,
it selects $R=\mathbf{84}$, as discussed in
Section~\ref{sec:bdstates}.

To study $\Pi_{\mathbf{84}}\ket{\mathcal O}$, we use the
coordinates of \eqref{eq:param} in the transverse gauge
\eqref{eq:gauge3pt} to parameterize the valley:
\begin{equation}
X^I=
\begin{pmatrix}
r \, \hat r^I/2 & (u^I+\mathrm i v^I)/\sqrt2\\
(u^I-\mathrm i v^I)/\sqrt2 & -r\, \hat r^I/2
\end{pmatrix},
\qquad
\hat r\cdot u=\hat r\cdot v=0.
\end{equation}
The coordinates $u,v$ describe off-diagonal fluctuations
about the commuting matrices $u=v=0$.
As described in Section~\ref{sec:3pt-LS}, the coordinates
$u,v$ in the transverse gauge \eqref{eq:gauge3pt} are
harmonic oscillators of frequency $r$.
Choosing $\hat r$ along direction $9$, define the creation operators
$a^\dagger_{u,i}$ and $a^\dagger_{v,i}$ with $i=1,\ldots,8$.
The combinations
$(a^\dagger_{u,i}\pm\mathrm i a^\dagger_{v,i})/\sqrt2$
carry charges $\pm1$ under the residual U(1) gauge
symmetry of the commuting background.
For either charge
there are eight bosonic and eight fermionic excitation
operators. Gauge invariance requires zero net charge,
so the first physical excited level contains an excitation with charge $+1$ and an excitation with charge $-1$. Thus its energy above the vacuum is $2r$,
and its $(8+8)^2=256$ states form one irreducible module
$S$ of the fast supercharges, as in
\eqref{eq:supermultiplet}.

We now determine which states in this oscillator multiplet
are created by $\mathcal O$. For SU(2), the antisymmetrized
product of three matrices allows us to write
\begin{equation}
\mathcal O\propto T^{IJK}B^{IJK},\qquad
T^{IJK}=\epsilon_{abc}X_a^I X_b^J X_c^K,\qquad
B^{IJK}=\frac{\mathrm i}{2}
\sum_a\psi_{\alpha a}\gamma^{IJK}_{\alpha\beta}
\psi_{\beta a}.
\end{equation}
Substituting \eqref{eq:param} gives
\begin{equation}
T=-\frac{r}{\sqrt2}\,\hat r\wedge u\wedge v
 =-\frac{1}{\sqrt2}\,
   \hat r\wedge\tilde u\wedge\tilde v,
\qquad
\tilde u=\sqrt r\,u,\quad \tilde v=\sqrt r\,v,
\end{equation}
using the rescaled coordinates of
Section~\ref{sec:3pt-LS}.
Acting on the oscillator vacuum, each transverse
coordinate acts as a creation operator. Thus $T$ creates the
antisymmetric combination
\begin{equation}
\hat{r}_I T^{Iij}\ket{0_{\hat r}} \propto \ket{A^{ij}}
\equiv \left(
a^\dagger_{u,i}a^\dagger_{v,j}
-a^\dagger_{u,j}a^\dagger_{v,i}
\right)\ket{0_{\hat r}}.
\end{equation}
This state belongs to the first physical excited level.
Its antisymmetric transverse polarization lies in the
$\mathbf{84}$ component of $S$: the $\mathbf{44}$ contains
no antisymmetric two-tensor under the SO(8) rotations
fixing $\hat r$.

Now decompose $B=B_{\rm fast}+b$, where
\begin{equation}
b^{IJK}=\frac{\mathrm i}{2}
\theta_\alpha\gamma^{IJK}_{\alpha\beta}\theta_\beta
\end{equation}
acts only on the slow fermions. In the contraction with
$T^{IJK} B_\text{fast}^{IJK}$, the fast contribution annihilates
$\ket{0_{\hat r}}$. Using the leading ground-state tail
\eqref{eq:tailr}, we therefore obtain
\begin{equation}
\bra{\vec r}\mathcal O\ket{\Omega}
\propto r^{-9}\sum_{i<j}
\underbrace{\ket{A^{ij}(r,\hat r)}}_{\text{fast excitation}}
\otimes
\underbrace{
\hat r_I b^{Iij}\ket{s(\hat r)}
}_{\text{slow fermion}},
\qquad r\to\infty.
\end{equation}
Here $\ket{s(\hat r)}\propto
\hat r_I\hat r_J\ket{IJ}$ is the slow $\mathbf{44}$
polarization appearing in \eqref{eq:tailr}, with
$I,J=1,\ldots,9$.
The slow bilinear produces a nonzero state in the
slow-fermion $\mathbf{84}$. The fast and slow factors
thus both lie in $\mathbf{84}$ components, with their
indices contracted to form an overall SO(9) singlet.

To obtain infinitely many bound states, we modify the trial state $\ket\O$ to contain
rapid radial oscillations. Define
$\rho^2=\sum_I\Tr[(X^I)^2]$, and choose a smooth gauge- and
rotation-invariant wavepacket $F(X)$ supported in a bounded region
far along a commuting valley. Keeping this cutoff fixed, consider
\begin{equation}
    \ket{\Psi_k}
    =\Pi_{\mathbf{84}}e^{\mathrm i k\rho}F(X)\ket{\O},
\end{equation}
where $\Pi_{\mathbf{84}}$ is the projector onto the internal
$\mathbf{84}$ component defined above. Both the cutoff and the radial
phase preserve the singlet condition. 
At large radial momentum,
the supercharges act on the slow fermions.
The nonzero slow-fermion $\mathbf{84}$ component of the trial
state's tail therefore ensures that the projection survives:
\begin{equation} \label{eq:normk}
    \lim_{k\to\infty}\braket{\Psi_k}{\Psi_k}=c>0.
\end{equation}
Clearly for large $k$ the Hamiltonian will grow without bound due to the kinetic energy. More carefully, differentiating $e^{\i k \rho}$ gives
$H\ket{\phi_k}=(k^2/2)\ket{\phi_k}+\ket{\chi_k}$,
where $\ket{\phi_k}=e^{ik\rho}F\ket{\O}$
and $\braket{\chi_k}{\chi_k}=O(k^2)$.
Since $\Pi_{\mathbf{84}}$ commutes with $H$, the $\sim k^2/2$ survives projection. 
Using \eqref{eq:normk}, we obtain
\begin{equation}
\frac{\bra{\Psi_k}H\ket{\Psi_k}}
     {\braket{\Psi_k}{\Psi_k}}
=\frac{k^2}{2}+O(k).
\end{equation}

As argued above, the positive-energy spectrum in this protected singlet sector
consists solely of normalizable bound states. 
If their energies were
bounded above, every state in the sector would have a bounded
mean energy, contradicting this construction. So there must
exist infinitely many SO(9)-singlet bound states with
$R=\mathbf{84}$ and energies tending to infinity.

This argument may be readily modified for $R = \mathbf{128}$ instead of $R= \mathbf{84}$. For this it is convenient to consider the trial state
\begin{align}
\ket{\mathcal O_{128}} &=\mathcal O_{128} \ket\Omega,\\
\mathcal O_{128} &=
\Tr(X^J\psi_\alpha)\,
\gamma^I_{\alpha\beta}\,
\Tr([X^I, X^J]\psi_\beta),
\end{align}
We leave the modification of this proof for this case as an exercise to the readers and/or their agents.

\clearpage %
\bibliographystyle{JHEP}
\bibliography{MyBib}

\end{document}